\documentclass[fleqn,usenatbib]{mnras}

\usepackage{newtxtext,newtxmath}

\usepackage[T1]{fontenc}
\usepackage{ae,aecompl}

\usepackage{graphicx}	
\usepackage{amsmath}	

\usepackage{color}

\newcommand{\sigmaZ}{$\sigma_{z}$ }
\newcommand{\sigmaR}{$\sigma_{r}$ }

\newcommand{\sigmaZR}{$\sigma_{z}/\sigma_{r}$ }
\newcommand{\sigmaZPHI}{$\sigma_{z}/\sigma_{\phi}$}
\newcommand{\sigmaPHIR}{$\sigma_{\phi}/\sigma_{r}$}

\title[Local variations of the SVE]{Local variations of the Stellar Velocity Ellipsoid-I: the disc of galaxies in the Auriga simulations
}

\author[D. Walo-Mart\'in et al.]{Daniel Walo-Mart\'in$^{1,2}$\thanks{E-mail: dwalo@iac.es},
Isabel P\'erez$^{3,4}$,
Robert J.J. Grand$^{5}$,
Jes\'us Falc\'on-Barroso$^{1,2}$, \and
Francesca Pinna$^{6}$,
and Marie Martig$^{7}$
\\
$^{1}$Instituto de Astrof\'isica de Canarias, Calle V\'ia L\'actea s/n, E-38205 La Laguna, Tenerife, Spain\\
$^{2}$Departamento de Astrof\'isica, Universidad de La Laguna, Av. del Astrof\'isico Francisco S\'anchez s/n, E-38206, La Laguna, Tenerife, Spain\\
$^{3}$Departamento de F\'isica Te\'orica y del Cosmos, Universidad de Granada, Facultad de Ciencias (Edificio Mecenas), E-18071, Granada, Spain\\
$^{4}$Instituto Carlos I de F\'isica Te\'orica y Computaci\'on\\
$^{5}$Max-Planck-Institut für Astrophysik, Karl-Schwarzschild-Str 1, D-85748 Garching, Germany\\
$^{6}$Max-Planck-Institut für Astronomie, Konigstuhl 17, 69117, Heidelberg, Germany\\
$^{7}$Astrophysics Research Institute, Liverpool John Moores University, 146 Brownlow Hill, Liverpool L3 5RF, UK\\
}

\date{Accepted XXX. Received YYY; in original form ZZZ}

\pubyear{2020}

\begin{document}
\label{firstpage}
\pagerange{\pageref{firstpage}--\pageref{lastpage}}
\maketitle

\begin{abstract}
The connection between the Stellar Velocity Ellipsoid (SVE) and the dynamical evolution of galaxies has been a matter of debate in the last years and there is no clear consensus whether different heating agents (e.g. spiral arms, giant molecular clouds, bars and mergers) leave clear detectable signatures in the present day kinematics. Most of these results are based on a single and global SVE and have not taken into account that these agents do not necessarily equally affect all regions of the stellar disc.We study the 2D spatial distribution of the SVE across the stellar discs of Auriga galaxies, a set of high resolution magneto-hydrodynamical cosmological zoom-in simulations, to unveil the connection between local and global kinematic properties in the disc region.
We find very similar, global, \sigmaZR = 0.80$\pm$ 0.08 values for galaxies of different Hubble types. This shows that the global properties of the SVE at z=0 are not a good indicator of the heating and cooling events experienced by galaxies. We also find that similar \sigmaZR radial profiles are obtained through different combinations of \sigmaZ and \sigmaR trends: at a local level, the vertical and radial components can evolve differently, leading to similar \sigmaZR profiles at z=0. By contrast, the 2D spatial distribution of the SVE varies a lot more from galaxy to galaxy. Present day features in the SVE spatial distribution may be associated with specific interactions such as fly-by encounters or the accretion of low mass satellites even in the cases when the global SVE is not affected. The stellar populations decomposition reveals that young stellar populations present colder and less isotropic SVEs and more complex 2D distributions than their older and hotter counterparts.

.

\end{abstract}




\section{Introduction}
\label{sec:Intro}





Late-type galaxies are systems dominated by rotation where most of the stars in the disc are expected to rotate in near circular orbits (epicycle approximation). Nevertheless, the presence of different structures is expected to modify the orbit of stars and leave distinct. For example, features such as a double thin-thick disc structure \citep{1979ApJ...234..829B,1979ApJ...234..842T}, warps \citep{1998A&A...337....9R,2006NewA...11..293A,2016MNRAS.461.4233R}, rings \citep{1986MNRAS.222..673F}, single and double bars \citep{1989Natur.338...45S,2019MNRAS.484..665D}, disc breaks \citep{1979A&AS...38...15V,2004ASSL..319..713P,2012MNRAS.427.1102M} and spiral arms \citep{1942AnLun..10..115D,2020IAUS..353..140B} leave different imprints in the kinematics and stellar populations properties that help us disentangle different evolutionary paths \citep{2019A&A...625A..95P,2020A&A...643A..14G,2020A&A...643A..65B,2020A&A...644A.116R}.

This type of analysis is extremely useful in the Milky Way where we have access to the full 6D information (positions and velocities), chemical abundances and ages of the stars. Thanks to missions like GAIA \citep{2016A&A...595A...1G} and APOGEE \citep{2017AJ....154...94M} we now have access to the properties of millions of stars in a large region of our Galaxy which allows us to study the spatial variations of the stellar kinematics beyond the solar neighborhood. 

The kinematics of stars within a volume can be characterized  by the three-dimensional distribution of velocities through the Stellar Velocity Ellipsoid (SVE). In a cylindrical rest frame, the SVE is represented by the velocity dispersion tensor whose diagonal elements and their ratios determine the level of random motions in the vertical, $\sigma_{z}$, radial, $\sigma_{r}$, and azimuthal directions, $\sigma_{\phi}$. Their ratio determine the shape of the ellipsoidal distribution. 
For instance, the spatial variations of the individual components and the shape of the SVE across the disc of our Galaxy for different stellar populations have been used to determine the existence of different heating mechanisms in the inner and outer parts \citep{2019MNRAS.489..176M}. On the other hand, the off-diagonal terms $\sigma_{zr}$, $\sigma_{z\phi}$ and $\sigma_{\phi r}$ determine the inclination of the velocity ellipsoid which is crucial to determine the distribution of baryonic and dark matter in the Galaxy. 
In particular, the inclination of the SVE in the ZR plane, known as the tilt angle, is important to properly determine the dark matter distribution in the galactic plane \citep{2015MNRAS.452..956B, 2019A&A...629A..70H,2019MNRAS.489..910E}. Furthermore, the inclination on the radial and azimuthal plane, so called vertex deviation, in the inner regions of our galaxy allows to study the presence of non-axisymmetries such as triaxial bulges or bars \citep{1994AJ....108.2154Z,2007ApJ...665L..31S,2020arXiv201113905S}.

Unfortunately, in external galaxies we do not have access to the properties of individual stars and it is difficult to perform a similar analysis. In the case of kinematics we only have access to the line-of-sight-velocity-distribution (LOSVD) which can be studied through parametric \citep{1993ApJ...407..525V,2011MNRAS.414..888E, 2017ApJ...835..104V} and non parametric techniques \citep[e.g.,][]{2020arXiv201112023F}. Therefore, to obtain the SVE parameters it is generally assumed that galaxies are in hydrostatic equilibrium and isothermal in the vertical direction \citep{1981A&A....95..105V,1988A&A...192..117V}, in the epicycle approximation, the individual velocity dispersions components follow an exponential distribution \citep{2003AJ....126.2707S} and that the axial ratios of the SVE are constant across the disc.

It is true that no galaxy is completely isolated and in steady-state, nor have a simple axisymmetric potential, but without these kind of assumptions it would not be possible to estimate the vertical velocity dispersion, $\sigma_{z}$, even in low inclination galaxies where \sigmaZ and the line of sight velocity dispersions are very similar \citep{2013A&A...557A.130M}. Furthermore, even though these assumptions simplify the complexity of these systems, with this type of approach we can still learn important information such as the mass of supermassive black holes \citep{2013ARA&A..51..511K} and dark matter haloes \citep{2014RvMP...86...47C}  using approximated dynamical models .


Under these assumptions \cite{2012MNRAS.423.2726G} studied the SVE of mildly inclined galaxies with different Hubble types and found a tight correlation between the morphological type and the $\sigma_{z}/\sigma_{r}$, where earlier type galaxies exhibited more isotropic ellipsoids.  On the contrary, \cite{PinnaSVE} (P18, hereafter) compiled the SVE axial ratios for a larger sample from the literature and found that this relation no longer holds. This result was further supported by the analysis of late-type simulated galaxies where the \sigmaZR ratio did not follow any obvious trend with the morphological type. Further, they studied the time evolution of the SVE and revealed that heating and "cooling" agents act at the same time and therefore the present day value of the SVE shape is not enough to infer the dynamical evolution of the galaxy. 

Thus, it is clear that the SVE of external galaxies is still not well known and we need higher spectral resolution observations to increase the quality and statistics of these analyses. On the other hand, theoretical studies on the SVE properties are required to provide testable predictions for these upcoming observations. To that end, cosmological zoom in simulations of late-type galaxies such as Auriga \citep{AurigaProject}, NIHAO-UHD \citep{2020MNRAS.491.3461B} and FIRE \footnote{https://fire.northwestern.edu} are extremely useful to study the evolution of different stellar properties across the galactic disc in a cosmological environment mimicking the results from coming observations.
 
In this work we aim to study the SVE of Milky-Way mass galaxies and in particular its local variation across the disc. We will focus on the Auriga galaxies, a set 30 zoom-in magneto-hydrodynamical cosmological simulations of late-type galaxies with different morphologies and evolution histories. These simulations successfully reproduce many present day observables such as stellar masses, sizes, rotation curves, star formation rates, and metallicities \citep{AurigaProject}. In addition, these galaxies present a variety of morphological features such as bars of different lengths, breaks \citep{2020MNRAS.491.1800B}, pseudo-bulges \citep{2019MNRAS.489.5742G} and warps \citep{2017MNRAS.465.3446G}. Thus, the Auriga sample is the ideal testbed because it will allow us to analyze the impact of external agents and different secular evolution processes in the SVE of high resolution disc galaxies in a cosmological environment. 

This paper is organized as follows. In Sec. \ref{sec:Auriga} we present the Auriga simulations, the sample selection and the methodology used to analyze the SVE. In Sec \ref{sec:Hubble_type} we study the connection between the global values of \sigmaZR and the morphology of galaxies. In Sec.\ref{sec:SVE_disc_spatial} we show the spatial variation of \sigmaZR across the disc. In Sec. \ref{sec:Time_evolution} we analyze the time evolution of the SVE at both global and local level. Sec. \ref{sec:Stellar_populations} presents the variations of \sigmaZR for different populations. In Sec. \ref{sec:conclusions} we sum up our conclusions.

Throughout the text, we will prefix units of length with 'c' to denote comoving scales e.g. ckpc for comoving kiloparsecs. Units without prefix should be interpreted as physical scales.

\begin{table*}
	\centering
	\caption{Parameters of the Auriga galaxies properties at z=0. SVE parameters indicate the median and 1$\sigma$ scatter across the disc dominated region. The columns are:(1) Galaxy name, (2) Hubble-Type, (3) Radius where the disc starts to dominate, (4) Stellar mass, (5) Vertical velocity dispersion, (6) Radial velocity dispersion and (7) Vertical to radial velocity dispersion ratio and (8) SVE categories from Sec. \ref{sec:SVE_maps}.}
	\label{tab:AU_param}
	\begin{tabular}{lccccccr} 
		\hline
		$Name$ & Hubble Type& $R_{\rm{0,disc}}$ (kpc) & $M_{\*}(10^{10}M_{\odot})$ & \sigmaZ (km/s) & \sigmaR (km/s) & \sigmaZR & Category \\
		\hline
		AU1 & SBb & 4.07 & 2.75 & 58$\pm$8 & 76$\pm$10 & 0.77$\pm$0.08 & PR\\
		AU2 & SBc & 8.99 & 7.05& 45$\pm$6 & 65$\pm$7 & 0.69$\pm$0.06 & HIAR\\
		AU3 & Sb & 7.26 & 7.75 & 51$\pm$7 & 60$\pm$12 & 0.83$\pm$0.07 & IAR\\
		AU4 & Sbc & 3.93 & 7.10 & 112$\pm$12 & 125$\pm$8 & 0.90$\pm$0.07 & PR\\
		AU5 & SBb & 4.58 & 6.72& 62$\pm$7 & 80$\pm$7 & 0.78$\pm$0.06 & HIAR\\
		AU6 & SBbc & 5.43 & 4.75 & 50$\pm$6 & 66$\pm$6 & 0.76$\pm$0.06 & PR\\
		AU7 & SBb & 5.43 & 4.88& 75$\pm$8 & 106$\pm$9 & 0.71$\pm$0.05 & PR\\
		AU8 & Sc & 6.57 & 2.99& 53$\pm$6 & 71$\pm$8 & 0.76$\pm$0.06 & SAR\\
		AU9 & SBb & 6.45 & 6.10& 59$\pm$8 & 78$\pm$7 & 0.75$\pm$0.07 & AR\\
		AU10 & SBa & 6.45 & 5.94& 82$\pm$14 & 95$\pm$11 & 0.91$\pm$0.15 & AR\\
		AU12 & SBab & 3.73 & 6.01& 73$\pm$9 & 91$\pm$14 &0.82$\pm$0.10 & PR\\
		AU13 & SBa & 5.43 & 6.19 & 101$\pm$10 & 103$\pm$8 & 0.97$\pm$0.07 & PR\\
		AU14 & SBb & 5.09 & 10.39 & 95$\pm$8 & 98$\pm$ 8& 0.97$\pm$0.08 & PR\\
		AU15 & Sbc & 5.32 & 3.93 & 59$\pm$7 & 75$\pm$ 8& 0.79$\pm$0.06 & PR\\
		AU16 & Sc & 9.03 & 5.41 & 53$\pm$9 & 64$\pm$12 & 0.82$\pm$0.06 & PR\\
		AU17 & SBa & 5.43 & 7.61 & 69$\pm$7 & 97$\pm$5 & 0.72$\pm$0.08 & AR\\
		AU18 & SBb & 6.45 & 8.04 & 65$\pm$11 & 80$\pm$5 & 0.83$\pm$0.13 & AR\\
		AU19 & SBc & 4.81 & 5.32 & 69$\pm$9 & 98$\pm$9 & 0.71$\pm$0.05 & HIAR\\
		AU21 & SBb & 3.39 & 7.72 & 71$\pm$9 & 86$\pm$9 & 0.84$\pm$0.07 & PR\\
		AU22 & SBa & 6.28 & 6.02 & 85$\pm$6 & 91$\pm$7 & 0.94$\pm$0.07 & AR\\
		AU23 & SBbc & 9.5 & 9.02 & 64$\pm$8 & 87$\pm$7 & 0.74$\pm$0.07 & AR\\
		AU24 & SBc & 5.09 & 6.55 & 52$\pm$7 & 72$\pm$7 & 0.73$\pm$0.06 & HIAR\\
		AU25 & SBb & 4.41 & 3.14 & 41$\pm$7 & 57$\pm$9 & 0.74$\pm$0.07 & SAR\\
		AU26 & SBa & 5.43 & 10.97 & 87$\pm$12 & 115$\pm$6 & 0.76$\pm$0.10 & AR\\
		AU27 & SBbc & 5.77 & 9.61& 68$\pm$8 & 85$\pm$7 & 0.81$\pm$0.07 & HIAR\\
		AU28 & SBa & 7.46 & 10.45 &98 $\pm$13 & 116$\pm$13 & 0.87$\pm$0.06 & AR\\
		\hline
	\end{tabular}
\end{table*}

\section{Methodology}
\label{sec:Auriga}

\subsection{The Auriga simulations}

The Auriga project \citep{AurigaProject} consists of 30 zoom-in high resolution simulations of haloes taken from the largest volume Dark Matter Only simulation of EAGLE (L100N1504) presented in \cite{2015MNRAS.446..521S}. Simulations were performed with the N-body, magnetohydrodynamics (MHD) code AREPO \citep{2010ARA&A..48..391S}. We label each simulation by 'Au - N' with N ranging between 1 and 30. Simulations adopt the cosmological values taken from \cite{2014A&A...571A..16P} ($\Omega_{m}=0.307$, $\Omega_{b}=0.048$, $\Omega_{\Lambda}=0.693$ and $H_{0} = 100 \ \mathrm{h} \ km s^{-1} \ Mpc^{-1}$, where $\mathrm{h} = 0.6777$). Dark matter and star particles have a typical mass of $m_{DM}\sim 3 \times 10^{5} M_{\odot}$ and $m_{b}\sim 5 \times 10^{4} M_{\odot}$. In this work we show results from simulations with these mass resolutions, which correspond to a level-4 resolution \citep[see Table 2 in][to find the characteristics of different resolution levels]{AurigaProject} if not otherwise specified. The comoving softening length for star and dark matter particles is set to $ 500 \mathrm{h}^{-1} cpc$ before z=1, when the simulation adopts a physical gravitational softening length of 369 pc. Groups and subhaloes  were identified by the friend-of-friends \citep{1985ApJ...292..371D} and SUBFIND \citep{2001MNRAS.328..726S} algorithms, respectively, and merger trees were constructed with the LHaloTree algorithm described in \cite{2005Natur.435..629S}. The Auriga simulations include an updated version of the sub-grid modules used in \cite{2014MNRAS.437.1750M,2014MNRAS.442.3745M} to account for physical processes that act below the resolution limit of the simulation. We refer the reader to \cite{AurigaProject} for a complete description of the models and the modifications introduced. 

We do not include in our study AU\_11, AU\_20, AU\_29 and AU\_30 because they exhibit strong signs of interactions in their z=0 surface brightness maps. From the final sample of 26 galaxies almost three quarters are barred systems (20 of them) and their structural properties are presented in \cite{2020MNRAS.491.1800B}.

\subsection{SVE characterization}
\label{sec:SVE_method}
To obtain the SVE we first place the disc of the galaxy into the XY plane by rotating the angular momentum vector of stars younger than 3 Gyr within a sphere of 60 kpc of radius around the center of potential. This angular momentum is dominated by the young thin disc which is less affected by external agents and guarantees that the galaxy is properly rotated even when it is interacting with its satellites. Then, to get the face-on view, we project galaxies along the Z axis onto a square grid of 0.5 x 0.5 kpc pixels. The number of particles decreases as we get further away from the center and so the information obtained from the individual pixels loses statistical significance in the outer regions of the galaxy. To ensure that we have a minimum number of particles we use the Voronoi binning routine of \cite{2003MNRAS.342..345C} to combine cells so every parameter is inferred with a minimum of 500 particles. This number of particles is larger than the typical 100 particles requirement used by \cite{2018MNRAS.480.4636S} or \cite{2020MNRAS.494.5652W} even though the mass resolution of Auriga is 2 and 100 times better than the simulations analyzed in these works. 
Furthermore, at that spatial resolution  imposed by that binning, it is still possible to distinguish spiral arms even in the outer parts of the galaxies. When stellar particles are divided into different populations, we will use the term standard to denote quantities derived from all the stellar particles.


To characterize the SVE we use a cylindrical restframe and we calculate the vertical ($\sigma_{z}$) and radial ($\sigma_{r}$) velocity dispersions and the axial ratio $\sigma_{z}/\sigma_{r}$ in each bin. We notice that the complete analysis of the ellipsoid should include the azimuthal velocity dispersion, $\sigma_{\phi}$, and one of the remaining axial ratios \sigmaPHIR  or \sigmaZPHI. Since the number of observational works concerning the SVE variations is quite limited and in general is focused to the vertical and radial components we decided to restrict our analysis to these components \citep[see][for an extensive analysis of the three components of the global SVE]{PinnaSVE}.

In this work we also want to characterize the global SVE properties of the disc through the median $\sigma_{z}$, $\sigma_{r}$ and  $\sigma_{z}/\sigma_{r}$. To properly select the disc dominated region we use the length of the bar, $L_{\rm{bar}}$, measured by \cite{2020MNRAS.491.1800B} and the disc scale length, $R_{\rm{d}}$, from \citep{AurigaProject} as the beginning of the disc dominated region, for barred and unbarred systems respectively. We only include all the bins with V-band surface brightness, $\mu_{\rm{V}}$, under 25 mag/arcsec$^{2}$. Thus, we exclude the innermost parts which are affected by structures such as bars and bulges and the low surface brightness region in the outskirts with undesirable large bins.

\section{Hubble Type relation}
\label{sec:Hubble_type}


\begin{figure}
\centering
\includegraphics[width=0.5\textwidth]{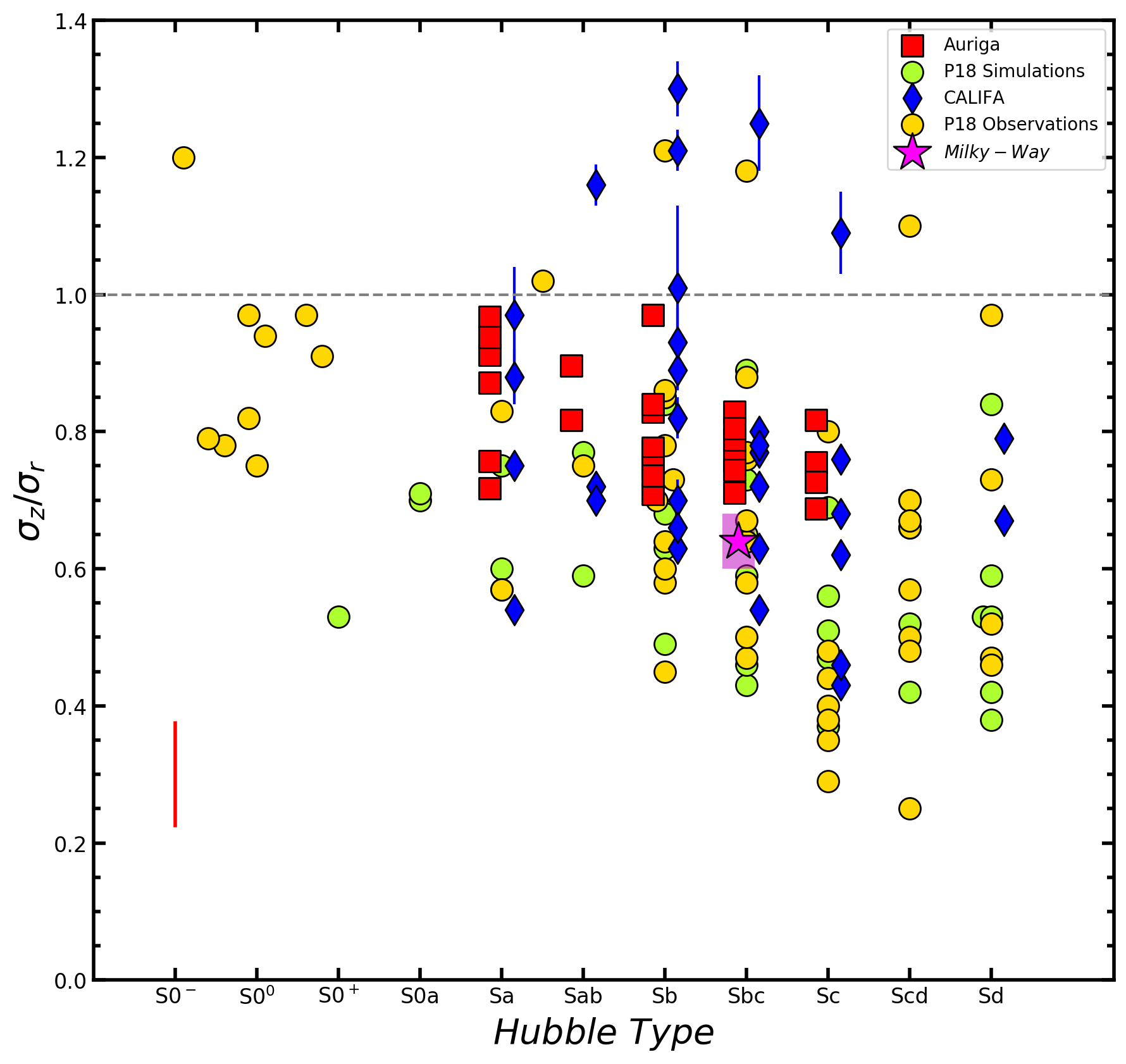}
\caption{Ratio of vertical to radial velocity dispersion as a functions of Hubble type. Yellow and light green circles represent the observed and simulated galaxies from P18. In blue, diamonds and error bars of galaxies in the CALIFA survey from \protect\cite{2019MNRAS.489.3797M}. 
Red squares indicate the median value of Auriga galaxies and the mean dispersion (84$^{\rm{th}}$-16$^{\rm{th}}$) in the sample is shown at the bottom left part of the diagram. The magenta star and shadowed region represent the Milky Way average value and dispersion from \protect\cite{2019MNRAS.489..176M}. The horizontal dashed line indicates the shape of an isotropic ellipsoid ($\sigma_{z} = \sigma_{r}$).}
\label{fig:SHAPE_HTYPE}
\end{figure}

In Sec. \ref{sec:Intro} we presented that there is a controversy about the connection between the shape of the SVE and the morphological type of a galaxy. The earlier results of \cite{2012MNRAS.423.2726G} showed that later type galaxies presented more oblate ellipsoids. On the other hand, P18 analysis of photometric observations \citep{1999A&A...352..129V}, spectroscopic decomposition of the line of sight \citep{2007MNRAS.379..418C,1999MNRAS.303..495E,2006MNRAS.371.1269T,2016MNRAS.460.2720K} and dynamical models \citep{2007MNRAS.379..418C,1999MNRAS.303..495E,2006MNRAS.371.1269T,2016MNRAS.460.2720K} did not exhibit such relation. In addition, P18 found that the simulations from \cite{2012ApJ...756...26M} and \cite{2017MNRAS.464.1502M} do not favor a scenario where \sigmaZR is connected with morphology. We refer the reader to Sec. 2 and 4 of P18 for a detailed analysis of each observational dataset and the adequacy of the different simulations. 
 As far as we know, this connection has not been further explored from the observational or theoretical point of view. Thus, we now study the SVE of Auriga galaxies to investigate this aspect in an independent dataset of simulations. The Hubble type of barred galaxies is based on the morphological classification of \cite{2020MNRAS.491.1800B} (see Table 1 in their work) while the six remaining unbarred galaxies have been visually classified in this work.
\begin{figure*}
\centering
\includegraphics[width=0.99\textwidth]{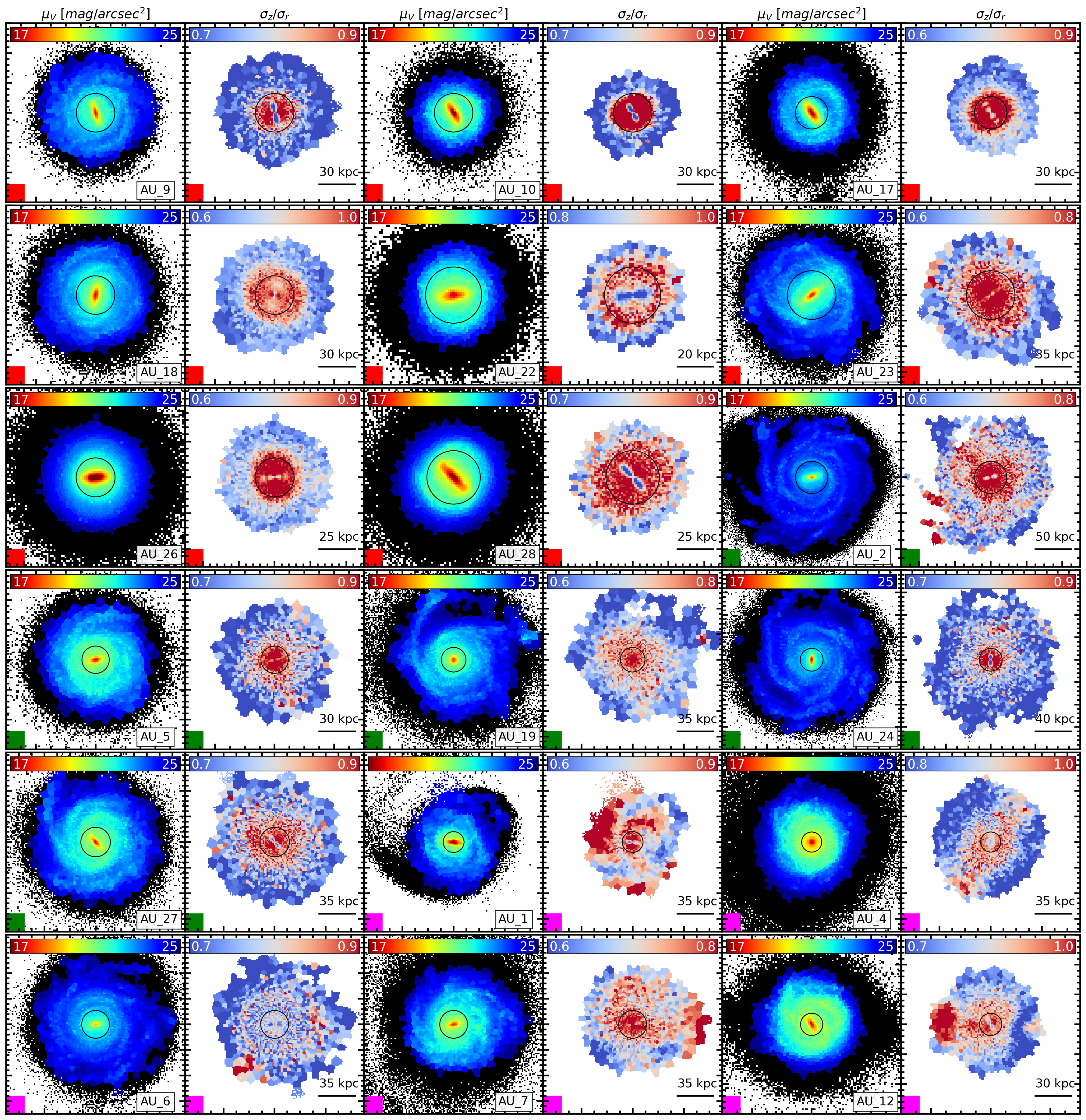}
\caption{V-band surface brightness (left) and \sigmaZR maps (right) for each of the galaxies in the Auriga sample. In the left panel, the bottom right box shows the galaxy name and the color scale ranges from 17 to 25 mag/arcsec$^{\rm{2}}$. Black bins indicate the regions with surface brightness $\mu_{\rm{V}}$>25. In the right panels, the color scale ranges between the 10$^{\rm{th}}$ and 90$^{\rm{th}}$ percentile of values in the disc region, and the box size is indicated in the bottom right part of the panel. Black circles represent the beginning of the disc region. Coloured boxes in the bottom left corner of each panel indicate the \sigmaZR category: AR (red), AIR (green), PR (magenta), SA (cyan), IAR (yellow).}
\label{fig:FIG_MAPS_ALL_1}
\end{figure*}


\begin{figure*}
\centering
\includegraphics[width=0.99\textwidth]{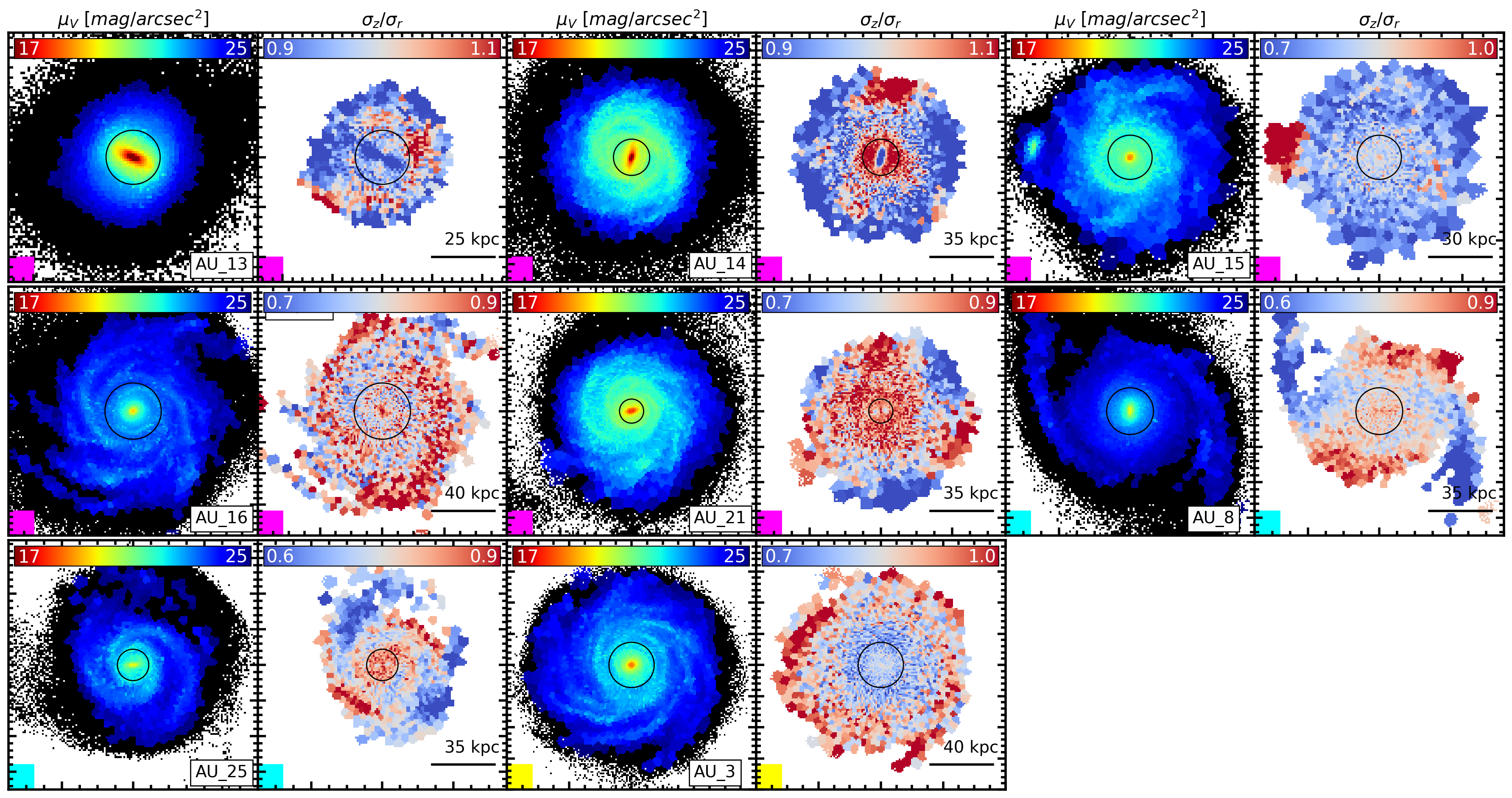}
\caption{Same as Figure 2}
\label{fig:FIG_MAPS_ALL_2}
\end{figure*}

In Fig. \ref{fig:SHAPE_HTYPE} we show the \sigmaZR values of galaxies as a function of their morphological Hubble type. Different symbols and colours have been used according to the source and type of the data. Red squares represent galaxies in the Auriga sample, with median \sigmaZR values in the 0.71-0.98 range and morphologies from Sa to Sc (see Table \ref{tab:AU_param} for the individuals values in the sample). The red vertical line in the bottom left part of the diagram indicates the average \sigmaZR scatter between the different bins (0.15) in the disc region for our sample. Yellow and light green circles show the axial ratios for the simulations and observations presented in P18. Both datasets contain galaxies with earlier and later morphologies than in Auriga. This observational dataset presents the smallest  SVE axial ratio values which close to the critical value above which galaxy discs are predicted to be stable against bending perturbations: \sigmaZR ~ 0.3 \citep{1964ApJ...139.1217T}. The magenta star and shaded region represent the Milky Way values from \cite{2019MNRAS.489..176M} with \sigmaZR = 0.64 $\pm$0.04. Furthermore, we represent with blue diamonds the \sigmaZR values for 34 galaxies from the CALIFA survey  \citep[][,see their Table 2]{2012A&A...538A...8S}. These values were obtained from \cite{2019MNRAS.489.3797M}, who analyzed the radial profiles of the radial velocity dispersions for 34 galaxies.  \cite{2019MNRAS.489.3797M} derived these values were from the anisotropy parameters $\beta_{z}$=1-$\sigma_{z}^{2}/\sigma_{r}^{2}$ obtained by \cite{2017MNRAS.469.2539K} using Jeans modeling. We did not include the results for NGC 5614, UGC 04132 and UGC 09892 because they present too large uncertainties. Some observed galaxies present \sigmaZR values above unity, which can be naturally produced by galaxy mergers and satellite accretion (see Fig. \ref{fig:FIG_5_TIME_GLOBAL} and the analysis in Sec.\ref{sec:Time_evolution} for a detailed analysis).

We first notice that there is no strong correlation between the \sigmaZR ratio and the morphology of Auriga galaxies, in agreement with P18. Nevertheless, Auriga and P18 datasets (both observational and simulated) present a moderate Spearman correlation coefficient around -0.45 and p-values below 1\%. We have further explored the relation between \sigmaZR and other more quantitative proxies of galaxy morphology e.g. the gas fraction, stellar mass and disc mass fraction (D/T) but both correlation coefficients and p-values showed significantly weaker connections. On the other hand, we find that there is a small difference between the simulated samples. Auriga galaxies present larger \sigmaZR values with an average value of 0.80 $\pm$ 0.08 while simulated galaxies from P18 have an average \sigmaZR of 0.60 $\pm$ 0.15. We have confirmed that this apparent discrepancy is not caused by differences in the stellar mass distribution of the samples. In particular, for galaxies with morphologies between Sa and Sd the distributions are very similar with average values of 4.8 $\times$ $10^{10} M_{\odot}$ in P18 and 6.6 $\times$ $10^{10} M_{\odot}$ in our sample, which precludes that our results are biased due to the mass function of Auriga. On the other hand, the aim of this work is not to determine the difference in the ellipsoid of different sets of simulations, but to show the existence or not of a correlation between the SVE and the morphology of a galaxy. It is no surprise that the results differ slightly given that each set of simulations use their own subgrid modules to implement the physics that take places at scales below the spatial resolution of the simulation and that they use different algorithms e.g. Auriga uses AREPO while \cite{2012ApJ...756...26M} use a Particle-Mesh code with gas dynamics modelled with a sticky-particle algorithm. We note that the precise values of \sigmaZR are likely affected by many different effects, including numerical heating \citep[e.g.][]{2013ApJ...769L..24S,2013MNRAS.434.2373R}and details of the subgrid ISM/feedback model. Determining the role of each of these effects is beyond the scope of this paper, which we defer to a dedicated future study.

In a similar way, CALIFA galaxies do not exhibit any correlation with the SVE axial ratio and the Hubble type, with Spearman correlation coefficient of -0.27 and p-value of 14\%). This result is particularly relevant because the sample from \cite{2019MNRAS.489.3797M} represents one of the largest datasets analyzed in an homogeneous way. Furthermore, it is an additional way to confirm that the combination of results from different methodologies in P18 observational data was not the main driver for the conclusions obtained. 
Both observational datasets reveal that, Auriga galaxies lack oblate ellipsoids with values between 0.4-0.6 and prolate ellipsoids with \sigmaZR >1 at all morphologies. Interestingly, If we combine all the datasets we obtain  a moderate correlation with Spearman coefficient of -0.45 and a p-value equal to 1.4 $10^{6}$\$, which differs from the strong correlation from \cite{2012MNRAS.423.2726G} with Spearman coefficient of -0.95 and p-value of  0.02\%.

In summary, future works on the analysis of large cosmological volume simulations such as EAGLE \citep{2015MNRAS.446..521S} or IllustrisTNG \citep{2018MNRAS.473.4077P} are necessary to properly understand the connection between the morphology of galaxies and the axial ratio of their SVE. Nevertheless, both simulations and observations agree that this connection is much weaker than previously stated.

\section{SVE across the disc}
\label{sec:SVE_disc_spatial}
\subsection{2D distributions}
\label{sec:SVE_maps}
Auriga galaxies have very similar global \sigmaZR with an average value around 0.8 but it is not clear if this value is a good proxy of the SVE axial ratio at a local scale or on the contrary, if it is just the result of combining high and low \sigmaZR regions. In this section, we explore how the SVE is distributed across the disc of the galaxies.

In Fig. \ref{fig:FIG_MAPS_ALL_1} and Fig. \ref{fig:FIG_MAPS_ALL_2} we plot for each galaxy a pair of maps: the V-band surface brightness, $\mu_{\rm{V}}$, in the left and the \sigmaZR map in the right. This is the first attempt in the literature to study the 2D spatial variations of the SVE. The color scale on the left panel ranges from 17-25 mag/arcsec$^{2}$ and bins with larger values are shown in black to identify tidal features in the outskirts of the galaxies. On the right panels we only show the \sigmaZR value for bins with $\mu_{V}$< 25 mag/arcsec$^{2}$. Smaller galaxies present smoother light distributions while larger galaxies exhibit features such as flocculent and grand design spiral arms. We first notice that there is not a simple and unique distribution of \sigmaZR ratios across the disc and therefore, the global shape of the ellipsoid is not a good indicator of its spatial distribution. 

We have visually divided our sample in six different categories in terms of the \sigmaZR spatial distribution in the disc dominated region. Although this classification scheme might seem arbitrary, it is not our intention to state that late-type galaxies should be classified into these categories, but rather emphasize the common features that we find among galaxies in the same group. 

\begin{figure*}
\centering
\includegraphics[width=0.99\textwidth]{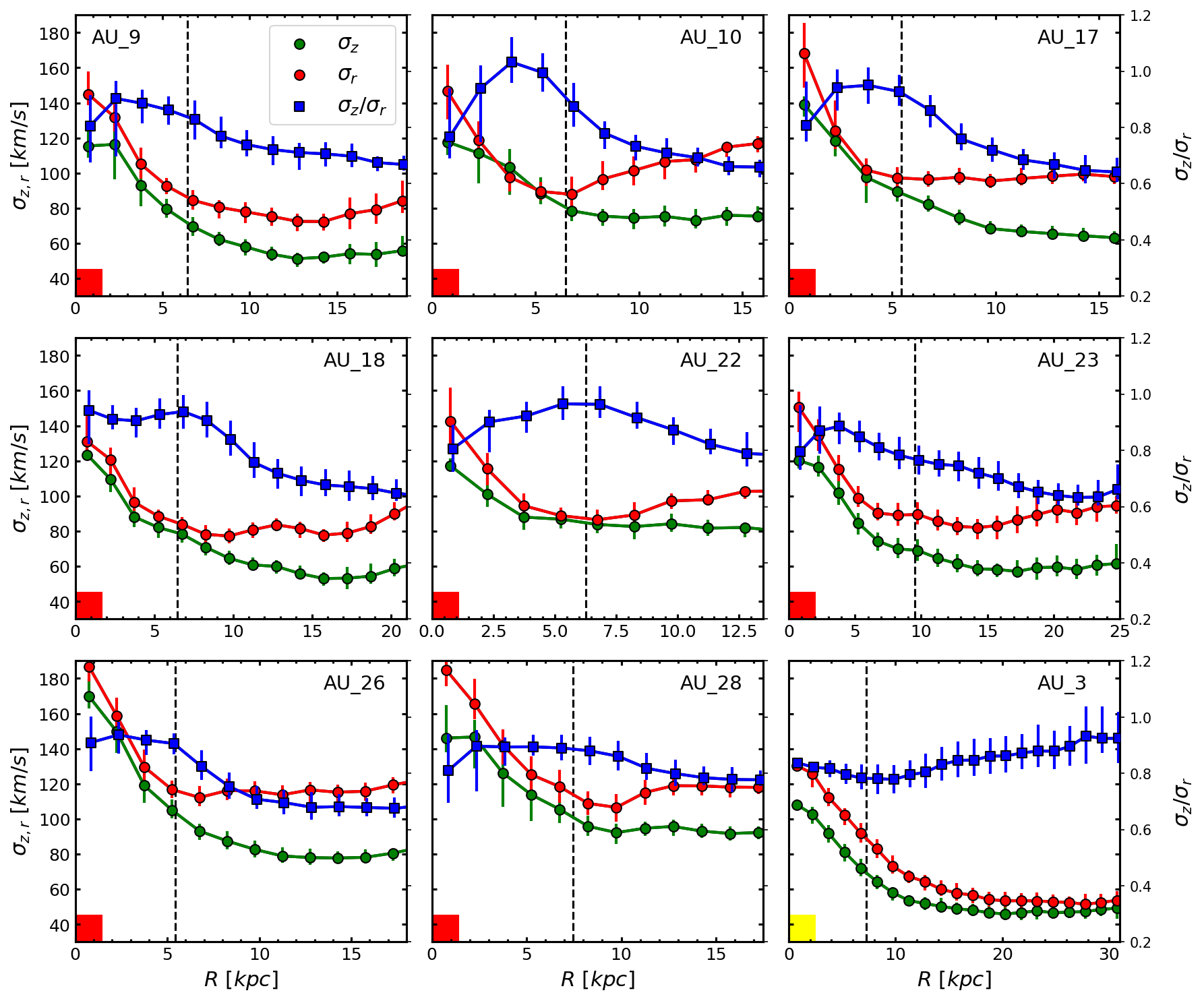}
\caption{Radial profiles of \sigmaZ (green circles), \sigmaR (red circles) and \sigmaZR (blue squares) in 2 kpc bins for galaxies with axisymmetric \sigmaZR distributions. The left and right vertical axis indicate the magnitude of the velocity dispersions and the axis ratio respectively. Symbols and errorbars represent the median, 16$^{\rm{th}}$ and 84$^{\rm{th}}$ value within each radial bin. The vertical dashed lines indicates the radius where the disc starts to dominate. }
\label{fig:FIG_4_PROFILES}
\end{figure*}

\textbf{Axisymmetric ratios (AR)}. The first group (red colour) is formed by galaxies with the most axisymmetric \sigmaZR distributions in the disc region. These systems present a clear transition from high \sigmaZR values in the inner parts to lower values in the disc region. All the members of this group are barred galaxies, which might suggest that the bar plays a stabilizing role in the disc dynamics. In fact, we find that the high to low \sigmaZR values transition occurs approximately at the length of the bar. Nonetheless, we note that our sample is biassed towards barred galaxies and we can not discard that this kind of distributions can also be obtained in unbarred systems. In this category we find AU\_9, AU\_10, AU\_17, AU\_18, AU\_22, AU\_23, AU\_26 and AU\_28. 

\textbf{High inner axisymmetric ratios (HIAR)}. The members of the second group (green colour) are galaxies with similar characteristics to those in the AR group (red colour), with higher values of \sigmaZR in the inner parts than in the outer disc and reasonable levels of axisymmetry. However, the transition region is not so well defined i.e. the inner high \sigmaZR regions expands beyond the length of the bar and the discs are less homogeneous with certain level of high \sigmaZR substructures. The former characteristics are best represented by the unbarred galaxy AU\_19, where the high \sigmaZR region expands up to twice the scalelength of the discs, which might, again, reflect the stabilizing role of the bar. In this category we also find AU\_2, AU\_5, AU\_24 and AU\_27.

\textbf{Perturbed ratio (PR)}. Following the previous two classifications from more to less homogeneous and axisymmetric \sigmaZR distributions, the third group (magenta colour) contains galaxies that present either high \sigmaZR features embedded in regular discs or entirely disturbed non-axisymmetric distributions. The former group is best represented by AU\_7, AU\_14 and AU\_15, where we see a rather regular \sigmaZR distribution in the disc region except for high value features in the outskirts. In the case of AU\_15 it is clear that this feature is caused by the nearby satellite, while for the others, interestingly, it does not seem to be a similar counterpart in the surface brightness maps. The rest of the galaxies, AU\_1, AU\_4, AU\_6, AU\_12, AU\_13, AU\_16 and AU\_21, present different levels of complexity in their \sigmaZR 2D maps and no special features in their surface brightness distribution. Nonetheless, we find some cases where the low surface brightness regions ($\mu_{\rm{V}}$ >25 mag/arcsec$^{2}$) present features that indicate that these systems are not in hydrostatic equilibrium, e.g. AU\_1 presents tidal features and AU\_4 exhibits an extended low surface brightness halo. 

\textbf{Spiral arm (SA)}.The next group (cyan colour) is formed by two galaxies, AU\_25 and AU\_8, that exhibit two symmetric spiral arms in the surface brightness maps with a clear imprint in the \sigmaZR maps. We notice that even though in the Auriga sample there are more galaxies with visible spiral arms (e.g. AU\_2 and AU\_16) typically they do not exhibit a clear \sigmaZR counterpart in the right panels. The panels with the cyan colour show a shared decrease \sigmaZR trend along the arm that indicates that the dynamical state of the stars is not the same within the spiral structure. The spiral arms of AU\_8 are particularly interesting because they exhibit the maximum \sigmaZR values in the inner parts of the arm and not in the central parts of the galaxy. The low surface brightness region of the galaxy shows that the arms are the most luminous part of a perturbed region. We will come back to this point in Sec \ref{sec:Time_evolution}.


\textbf{Inverted axisymmetric ratio (IAR)}. The last group (yellow color) is formed by AU\_3, the only galaxy that presents an inverted pattern with lower values of \sigmaZR in the inner parts and larger values in the outskirts. We notice that the galaxy presents an axisymmetric distribution but the peculiarity of this spatial distribution motivates its analysis as an independent group from the AR one.

We have additionally confirmed that there is no connection between the global \sigmaZR and the Hubble type for the AR, HIAR and PR categories by color coding each Auriga galaxy in \ref{fig:SHAPE_HTYPE} by their \sigmaZR classification (not shown). The plots for each of the groups showed that there is not dependence of \sigmaZR with the Hubble type of the galaxies, in a similar way to the full Auriga sample. Even though our classification in different groups is not physically motivated beyond the visual similitudes in the \sigmaZR maps, it is still relevant to remark that we obtain similar results within each subsample.

These results combined clearly expose that the global \sigmaZR of a galaxy is not a good indicator of the velocity structure in the disc because multiple spatial distributions of the SVE are compatible with the same global \sigmaZR. In fact, similar values do not only encompass galaxies with different radial trends but also systems with entirely perturbed patterns. This further indicates that the global ratio of the SVE is not a good indicator of the dynamical evolution of the galaxy.

\subsection{Radial profiles}
\label{sec:Radial_profiles}

Disturbed SVE patterns are a good indicator of galaxies that are not in hydrostatic equilibrium. Fly-by interactions and the accretion of satellites are expected to affect the disc unevenly, leading to different features in the \sigmaZR maps. On the contrary, the more regular \sigmaZR maps only indicate that the relation between the vertical and radial velocity dispersions is uniform across the disc, but is unclear whether they present similar spatial distributions of the individual components or different trends lead to the same $\sigma_{z}/\sigma_{r}$.

We now study in more detail the spatial variations of the individual SVE components, i.e. \sigmaZ and $\sigma_{r}$, to understand if galaxies with common 2D \sigmaZR maps also have similar velocity dispersions. To that aim, we obtain radial profiles of \sigmaZ and $\sigma_{r}$, and the \sigmaZR ratio in 1.5 kpc width radial bins. We limit this analysis to the galaxies with the most axisymmetric \sigmaZR maps (e.g. the AR and IAR groups) to ensure that within each radial bin we azimuthally average the regions with similar velocity structure. For example, the radial profile in the inner parts of barred galaxies do not capture the complex non axisymmetric features caused by the bar (see low \sigmaZR features along the bar in the maps of Fig. \ref{fig:FIG_MAPS_ALL_1} and Fig. \ref{fig:FIG_MAPS_ALL_2}).

Fig. \ref{fig:FIG_4_PROFILES} shows the radial profiles of \sigmaZ, \sigmaR and \sigmaZR for the 9 galaxies with the most axisymmetric distributions of $\sigma_{z}/\sigma_{r}$. Left and right vertical axis indicate the magnitude of the velocity dispersions and the SVE axis ratio, respectively. Green and red circles represent the vertical and horizontal velocity dispersions while the blue squares indicate the \sigmaZR values. Symbols and errorbars indicate the median, 16th and 84th percentile in each radial bin. The vertical line indicates the the beginning of the disc-dominated region.

The variety of \sigmaZ and \sigmaR radial profiles clearly show that the axial ratio of the SVE is not sufficient to comprehend the kinematic structure of the disc. The different panels reveal that in general, the individual velocity dispersions decrease outwards, but not always in an exponential way. Furthermore, in some cases the velocity dispersions flatten at a certain radius or even reverse their pattern showing larger values outwards. Galaxies like AU\_18 and AU\_23 present a more complicated pattern with small increases in \sigmaR at the position of the spiral arms seen in the surface brightness maps of Fig. \ref{fig:FIG_MAPS_ALL_1} around 13 and 10 kpc respectively. This in good agreement with the results from \cite{2019MNRAS.489.3797M}, where the authors found that $\sigma_{r}$ in late-type galaxies does not always decrease in an exponential manner. We notice nonetheless, that their analysis was limited to the inner parts within one effective radius and that in their methodology they assumed a constant \sigmaZR while we study the complete extension of the disc, and no assumptions on the axis ratio if the SVE is previously imposed. 
A more in depth analysis of the SVE profile of AR galaxies provides more information. First, we find that in general the SVE is more isotropic in the central parts than in the disc even though there are clear low \sigmaZR features along the bar in Fig. \ref{fig:FIG_MAPS_ALL_1} maps. Secondly, the disc dominated regions present a steadily decreasing \sigmaZR radial profile. In addition, \sigmaZ and \sigmaR show that this common behaviour of the SVE ratio is obtained through different combinations of \sigmaZ and \sigmaR. In particular, decreasing \sigmaZR profiles might be obtained by i) flat \sigmaZ and an increasing \sigmaR profile e.g. AU\_10 and AU\_22, or ii) flat \sigmaZ and a decreasing \sigmaR profiles e.g. AU\_17. Lastly, some galaxies present almost constant \sigmaZR in the outskirts of the disc, as in the case of AU\_26 and AU\_28. The cause of this behavior is that both the vertical and horizontal components become constant at certain radius which is the only way to obtain constant \sigmaZR in our sample.

AU\_3 is a galaxy with an atypical \sigmaZR distribution in both the inner and outer parts, in contrast to the rest of the axisymmetric galaxies. The central region shows a flat profile, which already indicates that \sigmaZ and \sigmaR present the same radial dependency, while the disc exhibits an increasing tendency in \sigmaZR outwards. This unusual behaviour could be interpreted as an increasing \sigmaZ trend caused by some dynamical heating event. However, the radial profiles of the vertical and radial components reveal that the axial ratio profile is due to a flattening of \sigmaZ at shorter radius than $\sigma_{r}$.

\begin{figure}
\centering
\includegraphics[width=0.5\textwidth]{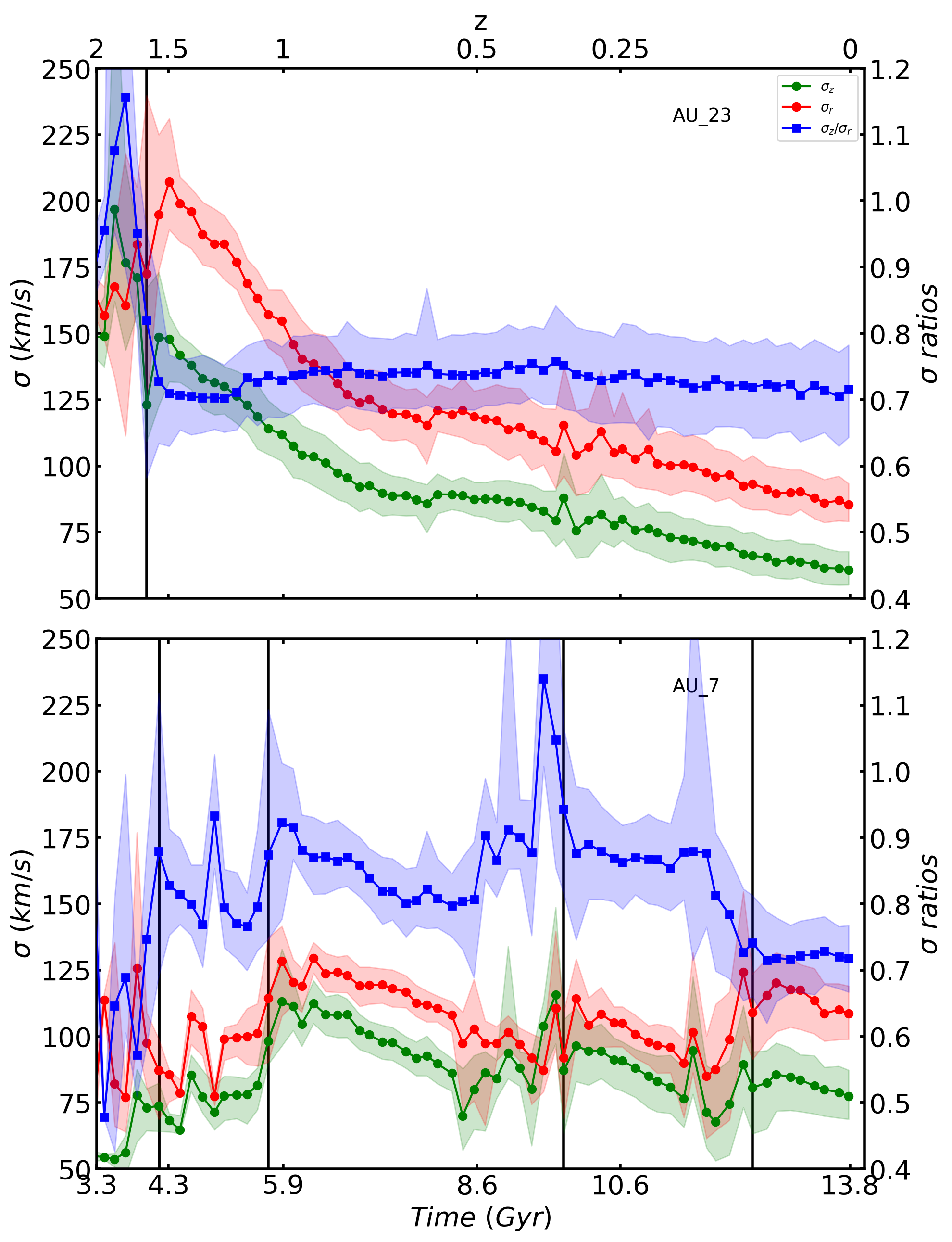}
\caption{Evolution of \sigmaZ (green circles), \sigmaR (red circles) and \sigmaZR (blue squares) as function of time since the Big Bang (bottom axis) and redshift (upper axis). From top to bottom we see the evolution of AU\_23 an isolated galaxy with no strong interaction and AU\_7, a galaxy that experiences multiple mergers. The left and right vertical axis indicate the magnitude of the velocity dispersions and \sigmaZR respectively. Symbols and the dashed area represent the median, 16$^{\rm{th}}$ and 84$^{\rm{th}}$ value of each parameter in the disc region. Vertical lines indicate mergers with mass ratio larger than 1:10}
\label{fig:FIG_5_TIME_GLOBAL}
\end{figure}

\begin{figure*}
\centering
\includegraphics[width=0.99\textwidth]{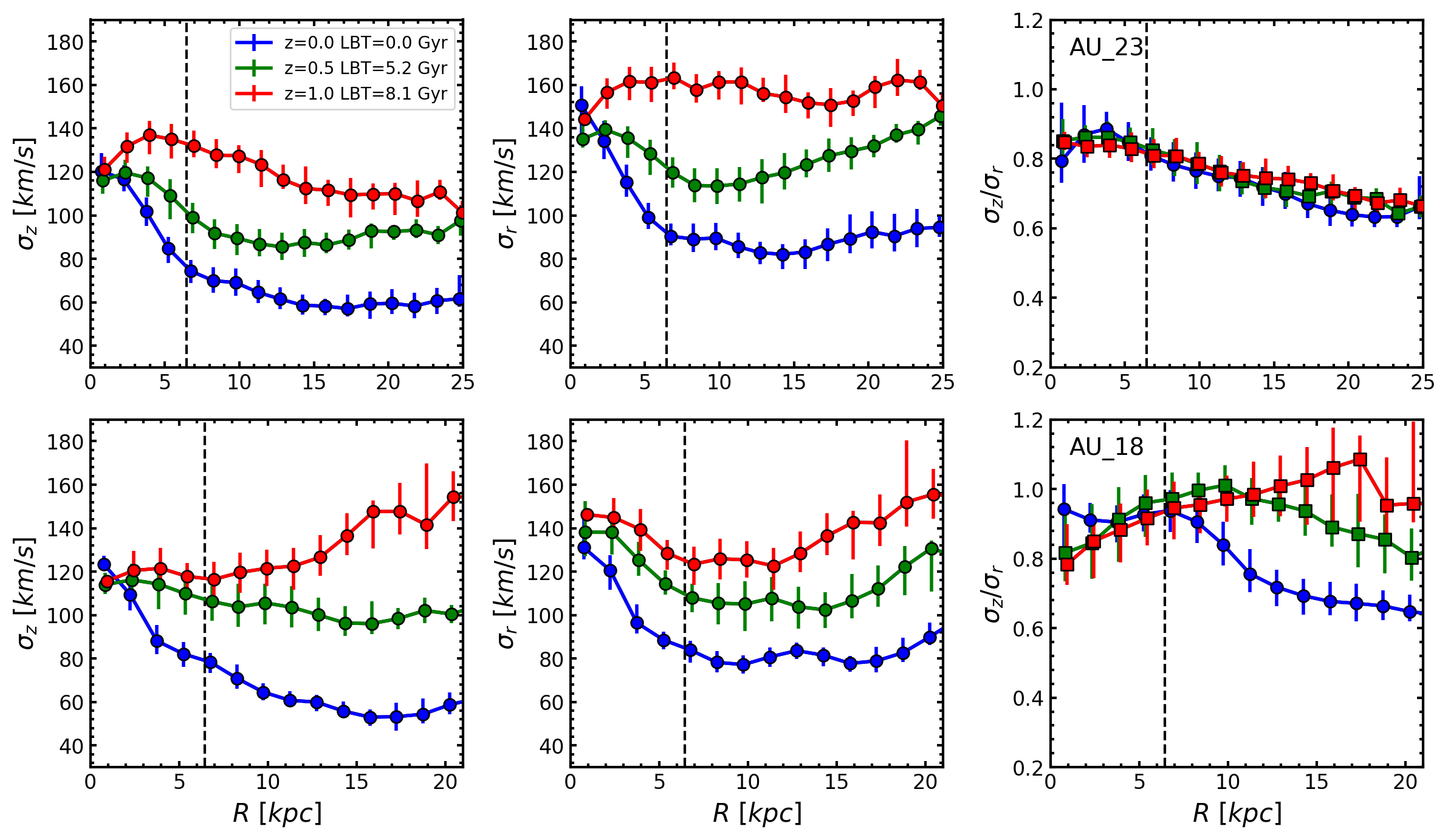}
\caption{Radial profiles of \sigmaZ (left), \sigmaR (center) and \sigmaZR (right) at z=0 (blue), z=0.5 (green) and z=1 (red) for Au\_23 (top) and AU\_18 (bottom). Symbols and errorbars represent the median, 16$^{\rm{th}}$ and 84$^{\rm{th}}$ value within each radial bin. The dashed vertical line represents the z=0 beginning of the disc dominated region.}
\label{fig:FIG_6_ExtraL}
\end{figure*}
 
To summarize, these results show that in general \sigmaZR is not constant within the disc region due to the complex combination of \sigmaZ and \sigmaR spatial distributions, except in the particular cases where both velocity dispersions have flattened. Our findings highlight the necessity of developing more elaborate models beyond the typically assumed exponential velocity dispersions and constant \sigmaZR to properly understand the dynamical state of real galaxies. For example, in \cite{2012MNRAS.423.2726G} it is assumed that \sigmaZ and \sigmaR have an exponential behavior with the same scale length and hence, the results can only provide a single \sigmaZR value for the entire disc. Nevertheless, there are a few works in the literature that have found that the \sigmaZR varies around 20\% within one effective radius in fast rotating early type galaxies e.g. \cite{2018MNRAS.477.3030K} and \cite{2008IAUS..245..215C}.

\section{SVE evolution}
\label{sec:Time_evolution}

Our results so far point to a scenario where galaxies have similar global values of \sigmaZR with no dependency on how this parameter is locally distributed or their morphology. This already illustrates the limited capability of this parameter to distinguish between different evolutionary scenarios. In addition, similar \sigmaZR distributions can be obtained through different combinations of \sigmaZ and \sigmaR, increasing the level of complexity when inferring the dynamic evolution of galaxies from this parameter alone.
In this section, we study the evolution of the SVE at both global and local levels to understand the role of different agents and up to what extent the SVE parameters preserve any information about them at z=0 .

\begin{figure}
\centering
\includegraphics[width=0.5\textwidth]{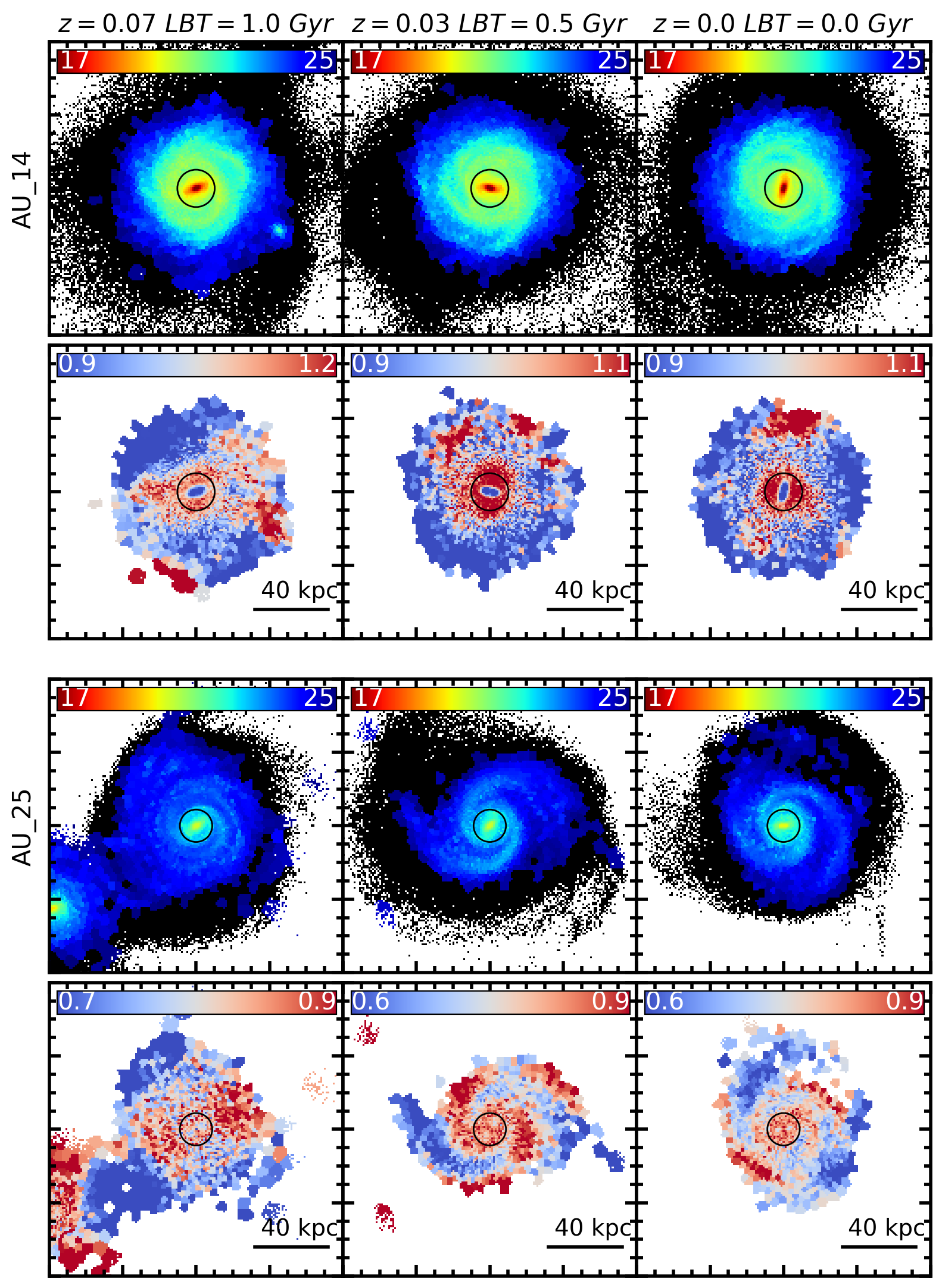}
\caption{V-band surface brightness (upper maps) and \sigmaZR maps (inferior maps) for AU\_14 (top panels), AU\_25 (bottom panels) at three different snapshots. Surface brightness colorscale ranges from 17 to 25 mag/arcsec$^2$ and bins with larger values are shown in black. Redshift and look-back time (LBT) are indicated on top. The box size of the field of view is indicated in the bottom right part of the panel. Black circles represent the beginning of the disc region.}
\label{fig:FIG_6_TIME_MAPS}
\end{figure}

\begin{figure}
\centering
\includegraphics[width=0.5\textwidth]{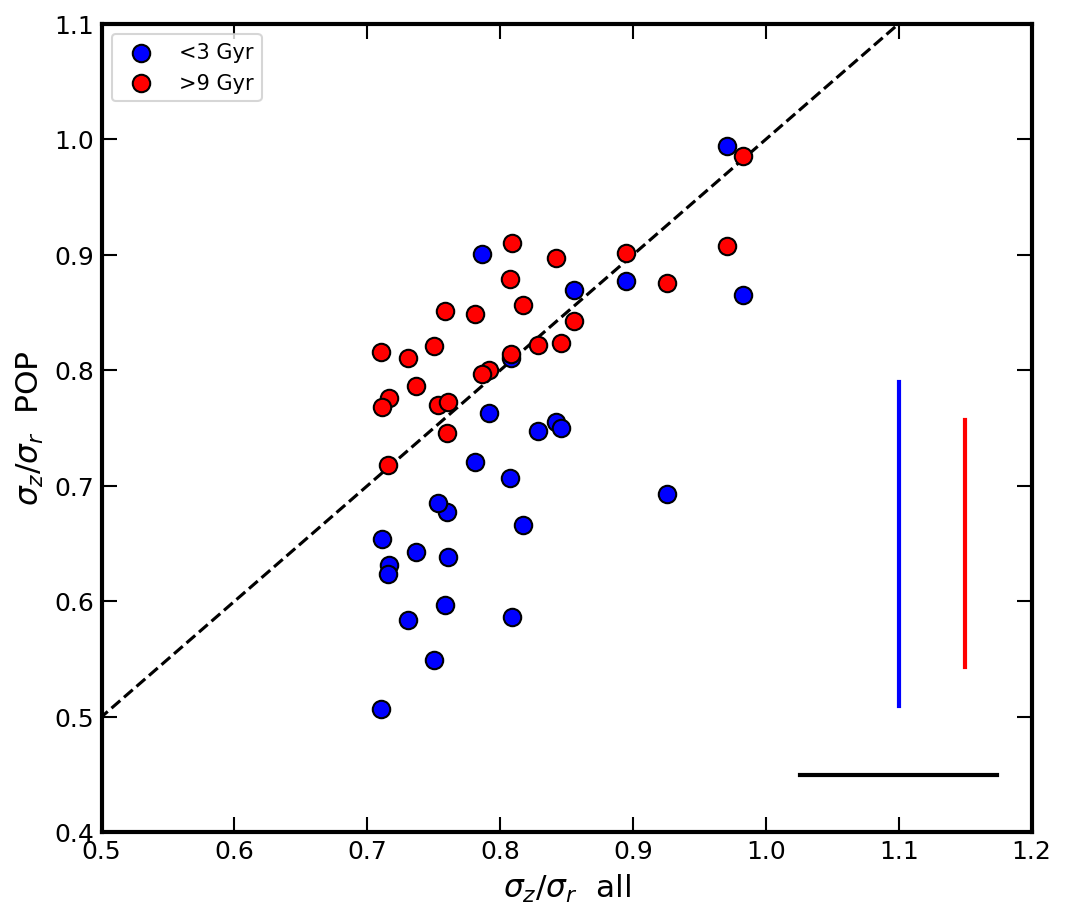}
\caption{Axis ratio of the SVE, $\sigma_{z}/\sigma_{r}$, for a stellar population as a function of the axis ratio obtained for all the stellar particles. Blue and red circles are used for the young (<3 Gyr) and old (<9 Gyr) populations comparisons. Vertical and horizontal lines at the bottom left part of the panel indicate the mean dispersion (84$^{\rm{th}}$-16$^{\rm{th}}$) in each dataset.}
\label{fig:FIG_ZR_POP_GLOBAL}
\end{figure}

Fig. \ref{fig:FIG_5_TIME_GLOBAL} shows the SVE evolution from z=2 of two galaxies with entirely different dynamical histories, AU\_23 (top) and AU\_7 (bottom). We use the same disc dominated physical region in different snapshots based on the z=0 photometric analysis. In the upper panel, AU\_23 evolves in a relatively isolated way with no significantly strong SVE variations for most of the time. In the time period between z=2 and z=1, both \sigmaZ and \sigmaR increase due to the accretion of two minor satellites and a larger merger soon after. These interactions do not equally affect both components, which leads to a maximum of \sigmaZR around 1.15 at z=1.75. Then, a rapid drop to 0.7 is caused by the last merger of the series which increases \sigmaR significantly. Then, after z=1.5, both velocity dispersions components decrease at similar rates, and hence the shape of the global SVE remains almost constant. The bottom panel shows a more complicated picture for AU\_7, where the galaxy experiences multiple episodes of heating caused by fly-by interactions and merger that have severely modified the global SVE at different times. 
It is particularly interesting to find that mergers do not always lead to more isotropic SVEs. For example, the last merger increases \sigmaR to a greater extent than $\sigma_{z}$, decreasing \sigmaZR even during a heating process. On the other hand, it is remarkable that every heating event is followed by a cooling period of time where the global vertical and radial velocity dispersions decrease. 

This "cooling" process is caused by the formation of stars in colder orbits which dominate the evolution of the global ellipsoid and should not be interpreted as a cooling episode of the preexisting populations. The decrease of the velocity dispersion of the gas over cosmic time has been confirmed by both observations \citep{2009ApJ...706.1364F,2014MNRAS.437.1070G,2015ApJ...799..209W} and cosmological simulations \citep{AurigaProject,2019MNRAS.490.3196P} and thus our result is expected. Nonetheless, it is remarkable that this effect is able to compensate for the heating caused by secular and external processes. For example, \cite{2016MNRAS.459..199G} showed by analyzing the evolution of a coeval population that the bar is one of the main source of heating of preexisting populations after it is formed. Here we show that the combination of all stellar populations lead to a steady decrease of the global SVE velocity dispersion. 

The rest of the Auriga sample shows an intermediate behaviour between the two extreme cases in Fig. \ref{fig:FIG_5_TIME_GLOBAL}. Galaxies combine secular evolution periods of time with varying number of heating events from mergers depending on the specific conditions of each system. When galaxies undergo one of these quiescent episodes \sigmaZ and \sigmaR might decrease at different rates, leading to increasing or decreasing \sigmaZR global values over time. These results are in excellent agreement with P18 which also revealed that the combination of heating and cooling mechanisms made it difficult to determine what processes had galaxies experienced (see Sec. 5.3 of their work for a detailed study on the effect of different agents in the evolution of the SVE).

Our methodology allows us to further study how the spatial distribution of the SVE changes over time. In particular, we investigate if the SVE radial trends are similar at all times while the global SVE evolves secularly.
In Fig. \ref{fig:FIG_6_ExtraL} we show, from left to right, the radial profiles of \sigmaZ, \sigmaR and \sigmaZR at z=0 (blue), z=0.5 (green) and z=1 (red) for AU\_23 (top) and AU\_18 (bottom). The latter galaxy has a dynamical history similar to AU\_23, evolving in isolation after a merger at z$\approx$1.25. However, from z=0.5, \sigmaZ cools at a faster rate than \sigmaR which leads to a slightly decreasing \sigmaZR trend in the last 5 Gyrs compared to the flat tendency of AU\_23 in Fig. \ref{fig:FIG_5_TIME_GLOBAL}. 

The radial profiles on the left and middle panels show larger values of \sigmaZ and \sigmaR in the disc region at earlier times, as expected from their global SVE evolutions. This is also true for some of the central regions except for the innermost parts where the velocity dispersions are very similar for the three curves. These plots show that the radial behaviour of \sigmaZ and \sigmaR is not constant over time, which further reveals that galaxies evolve in a more complicated manner than simply decreasing their global velocity dispersion since. Moreover, each component might evolve differently. For instance, we find that AU\_23 and AU\_18 experience more changes in \sigmaR and \sigmaZ profiles, respectively.

We find that \sigmaZR radial profiles evolve in entirely different fashions for each galaxy even though they shared some similarities in the evolution of \sigmaZ and $\sigma_{r}$. AU\_23 shows almost the same radial distribution of \sigmaZR at different snapshots which reveals that heating and cooling agents modify \sigmaZ and \sigmaR in a uniform way. On the contrary, AU\_18 presents larger values and smoother profile at z=0.5 and an increasing profile at z=1. The latter is clearly caused by the merger at z=1.25 as we can already infer from the increasing \sigmaZ profile. The large errorbars, indicating the scatter of averaged points at different azimuths, show that the interaction does not homogeneously affect the disc. A more detailed study of the mechanisms driving the evolution of the SVE at a local level for different galaxies exceeds the scope of this paper and will be addressed in future works. We remark, nonetheless, that present day axisymmetric \sigmaZR can be obtained through different processes.

We also study the evolution in time of the full 2D maps to investigate the origin of some of the features shown in Sec. \ref{sec:SVE_maps} that would otherwise be neglected in the radial profile. 
Fig. \ref{fig:FIG_6_TIME_MAPS} shows the V band surface brightness, $\mu_{\rm{V}}$ and \sigmaZR maps for two galaxies AU\_14 (top) and AU\_25 (bottom) at different snapshots at recent times (in a similar way to Fig.\ref{fig:FIG_MAPS_ALL_1}). 
From left to right the panels show the galaxies in three different snapshots, 1 Gyr ago, 0.5 Gyr ago and at present time. The global SVE parameters for these two galaxies follow a quiescent evolution that do not justify the formation of the \sigmaZR features seen in the maps.

In the top panels, we observe how the accretion of a low mass satellite, (1:30 mass ratio), induces the high \sigmaZR features in the z=0 map of AU\_14. In the left panel, the satellite is in the outskirts of the disc before it is finally accreted, leaving no trace of this event in the surface brightness map. This interaction increases the vertical velocity dispersion which causes the high \sigmaZR feature. The results are particularly relevant because they point to the possibility of using these maps to unveil the accretion of low mass satellites in future observations.

AU\_25 presents a different picture with very different maps both in the surface brightness and \sigmaZR in the three snapshots. 
The dramatic change from z=0.1 to z=0 is caused by the interaction with a massive satellite with mass ratio 1:3. In the left panel, we can see how the companion is approaching AU\_25. In the next panel, the apocentric passage of the satellite has induced the formation of two large symmetrical spiral arms that are also present in the \sigmaZR map. In the third panel we observe that the effect of the fly-by is still present in both maps at present time. 

The second galaxy with spiral patterns in the surface brightness and \sigmaZR maps, AU\_8, reveals a similar story. The galaxy presents a regular disc structure that is slowly setting into equilibrium after a minor merger 3.5 Gyr ago. Then, a fly-by interaction 0.7 Gyrs ago caused the formation of two symmetric spirals arms at z=0 (see Fig.\ref{fig:FIG_MAPS_ALL_2}). After a thorough inspections of our sample, we find that other galaxies have exhibit similar features but their signatures have disappeared. The reason why these patterns are present at z=0 is because they have recently occurred and galaxies have not been able to reach equilibrium.

In summary, the most interesting aspect of these interactions is that even though they do not affect the global SVE parameters they may leave clear imprints in the SVE local distribution.

\section{The effect of stellar populations on the SVE}
\label{sec:Stellar_populations}

Our work provide valuable insights from the theoretical point of view because the number of studies in the literature focused on the detailed analysis of the SVE is small. However, our methodology does not take into account observational selection effects. In particular, our analysis of the SVE is based on mass weighted quantities which is relevant as these parameters are more directly connected with the dynamic of the stars. 

It is expected that the SVE varies for different stellar populations since mergers take place at different times and the interstellar medium from which stars are formed steadily cools down. Nevertheless, it is unclear if the variations in the mass to light ratio for different populations might cause that the observational (and thus luminosity weighted) SVE deviates significantly from the mass weighted one. To account for this effect we have repeated our analysis weighting the contribution of each stellar particle to the SVE by their V-band luminosity. The spatial distribution of the three SVE parameters is very similar to the mass weighted but experience a 20\% decrease in \sigmaZ and \sigmaR while \sigmaZR is of the same order. 


The previous results show up to what extent young populations affect the luminosity weighted SVE when all the stellar particles are considered. However, the extent to which the global and local SVE agree with one another for different stellar populations is unclear. To answer this question we focus on a group of young (<3Gyr) and old (>9Gyr) stars and study their SVE within the same spatial bins we have used through this work. We notice that the SVE of young stars would be best described by a shorter age window since certain interaction induce rapid changes in less than 1 Gyr, but this limit ensures that we still have a statistically significant number of particles in each bin.
\begin{figure*}
\centering
\includegraphics[width=0.95\textwidth]{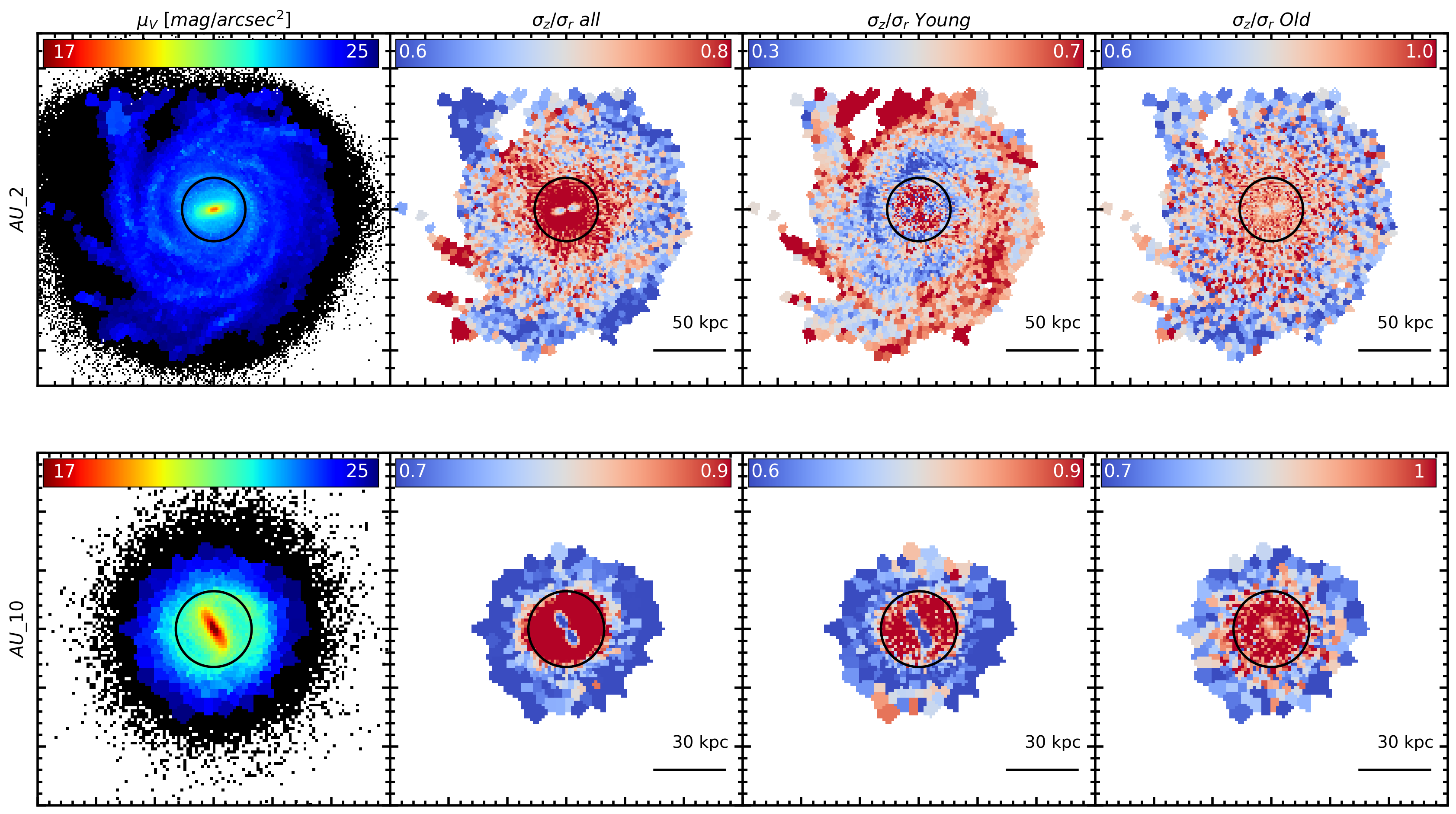}
\caption{From left to right we show the V-band surface brightness in the 17-25 mag/arcsec$^{2}$, \sigmaZR map of all the stellar particles, \sigmaZR map of stellar particles younger than 3 Gyr and \sigmaZR map of stellar particles older than 9 Gyr for AU\_2 (top) and AU\_10 (bottom).The upper left box in the \sigmaZR maps show the color scale range and the physical scale of the field of view is indicated in the bottom right part of the panel. Black circles represent the beginning of the disc region}
\label{fig:FIG_ZR_POP_MAPS}
\end{figure*}

We first found that the global \sigmaZ and \sigmaR of the young and old components are on average 50 \% cooler and hotter than the standard global SVE, which is in agreement with the global evolution results and the works previously mentioned  \citep[e.g.][]{AurigaProject,2019MNRAS.490.3196P}. On the other hand, \sigmaZR results are more complicated. Fig. \ref{fig:FIG_ZR_POP_GLOBAL} shows the global \sigmaZR of the young (blue) and old (red) SVE as a function of the standard SVE axial ratio. Lines in the bottom right represent the average \sigmaZR scatter across the disc in our sample for each type of SVE, e.g. 0.3 young (vertical blue), 0.2 old (vertical red) and 0.15 standard (horizontal black). The dashed line indicates the one-to-one correlation. The panel shows that the young populations presents a more oblate ellipsoid, whereas the old population presents an ellipsoid similar to the standard SVE. This difference is in good agreement with the results for the Milky Way from \cite{2019MNRAS.489..176M} where the authors found in the 6-10 kpc region of the disc that young and old populations exhibited \sigmaZR values $\approx$ 0.4 and $\approx$ 0.6, respectively. 

Following the structure from previous sections we study in Fig. \ref{fig:FIG_ZR_POP_MAPS} how the spatial distribution of the ellipsoid changes for different populations. From left to right we show the V-band surface brightness and the \sigmaZR of the standard, young and old SVEs of AU\_2 (top) and AU\_10 (bottom). We first notice that the differences on the global values of \sigmaZR for different populations from Fig. \ref{fig:FIG_ZR_POP_MAPS} do not always correspond to global differences across the disc. The third and fourth column clearly illustrate how the spatial distribution of the SVE changes for different populations.

In the top panels, AU\_2 shows that galaxies may present different 2D distributions depending on the population that is studied. The spiral-like features in the third panel are particularly important because they reveal that spiral arms have an impact in the SVE of the young population even though the standard SVE presents a more regular pattern. On the other hand, old populations have a smoother and more axisymmetric \sigmaZR distribution than the standard SVE. The panels motivate a more in depth analysis on the spiral imprints in the SVE maps for different populations. To this end we studied the difference between the standard, young and old SVE parameters in the arm region (traced by the bins with high star formation rate) and the underlying stellar disc. However, the results did not exhibit any significant difference between them. The reason for this outcome is that there is not a perfect match between the spiral features in the surface brightness and \sigmaZR maps causing that SVE features are shared between the arm and inter-arm photometric regions. This trend is consistent with streaming motions indicative of radial migration along spiral arms \citep{2016MNRAS.460L..94G,2016ApJ...830L..40S}, and emphasizes the complex impact of spiral arms in the dynamics of the disc. The bottom panels show an opposite picture, where the \sigmaZR spatial distribution of AU\_10 is very similar for the different stellar populations. The only small difference is that there are more region of the disc with higher \sigmaZR values.

We want to remark that AU\_2 and AU\_10 are two interesting cases in our sample but they should not be interpreted as the only two possible scenarios in which the SVE maps change. Auriga galaxies show a complex SVE-stellar population dependence that significantly changes from galaxy to galaxy depending on their dynamical history. However, the detailed characterization of the SVE for different populations and their evolution is beyond the scope of this work. Nonetheless, these results show that the SVE properties for different populations is a promising tool for unveiling the formation mechanisms of galaxies.

Finally, we note that stars can only be resolved in a handful of galaxies in the nearby volume, hence our results may not be easy to confirm from an observational point of view. However, the latest advance in dynamical models such as the orbit-based Schwarzschild method \citep{2008MNRAS.385..647V,2008MNRAS.385..614V,2020ApJ...889...39V} allows to infer the underlying orbital distribution of stars within a galaxy without any ad-hoc assumption. In particular, the latest population-orbit method by \cite{2020MNRAS.496.1579Z} is a promising technique that has successfully incorporated the age and metallicity information into the orbit decomposition. Thus, observational analysis similar to that presented here will be possible in the near future.

\section{Conclusions }
\label{sec:conclusions}
We have analyzed a set of 26 high resolution cosmological zoom-in simulations of Milky-Way mass late-type galaxies from the Auriga sample and we have obtained maps of vertical and radial velocity dispersions and their ratio (e.g. \sigmaZ, \sigmaR and \sigmaZR respectively), to study the spatial variations of the Stellar Velocity Ellipsoid (SVE).

Following P18, the galaxies in our sample exhibit very similar \sigmaZR ratios and we do not find the connection with morphological type shown by \cite{2012MNRAS.423.2726G}. Auriga galaxies present more isotrope SVEs than P18 with a mean \sigmaZR value of 0.80 $\pm$ 0.08. We have also shown that this relation is not present in more recent observational studies using galaxies in the CALIFA survey.

Although galaxies present similar values of \sigmaZR the 2D spatial distribution of the SVE reveals different behaviours. In general, galaxies exhibit decreasing \sigmaZR outwards, except for AU\_3 which is the only case in our sample with an increasing \sigmaZR profile. The different levels of axisymmetry in the disc region motivated the classification of our galaxies into different categories from more to less homegeneous \sigmaZR distributions. Interestingly, we found that galaxies may present high \sigmaZR features in their map that are not associated with any visible feature in their surface brightness maps. On the other hand, our results show that even though many of the galaxies present spiral arms, only two of them exhibit clear features in their \sigmaZR maps. In these peculiar systems the spiral arms are triggered by recent interactions.

We have analyzed in more detail the radial variation of \sigmaZ, \sigmaR and \sigmaZR of galaxies that presented the most axisymmetric \sigmaZR maps. In general, the vertical and radial velocity dispersions decrease outwards but, in some cases, these profiles flatten or even increase after reaching a minimum value. Radial profiles of \sigmaZR tend to decrease outwards but similar behaviours can be obtained with different combinations of \sigmaZ and \sigmaR trends. In particular, we find that the combination of flat \sigmaZ with increasing \sigmaR and flat \sigmaR with decreasing \sigmaZ lead to the declining \sigmaZR profile. On the other hand, if each component decreases differently the resulting radial \sigmaZR will exhibit either an increasing or decreasing \sigmaZR profile. This further emphasizes that the \sigmaZR parameter is not able to capture the complex kinematic structure of galaxies.

The time evolution of the global SVE shows that the global \sigmaZR is not useful to determine the dynamical evolution of the galaxy since similar axial ratios can be obtained through different mechanisms. In the absence of mergers, the evolution of the global SVE is determined by the formation of stars in colder orbits which cause \sigmaZ and \sigmaR to steadily decrease with time. Simultaneously, the SVE ratio reaches a steady value fairly early then does not evolve much. Although this gives the impression of quiescent evolution, the spatial distribution of \sigmaZ and \sigmaR may vary over time and exhibit different radial trends. We find that, in some cases, the changes in the profiles of \sigmaZ and \sigmaR take place in a similar way throughout the disc region, preserving the same \sigmaZR distribution at all times, or in a more complex way, leading to different radial profiles at earlier epochs.

Further, we have shown that the evolution of the full 2D spatial distribution can be used to unveil the origin of some of the features seen at the z=0 maps. We found evidence that, in some cases, the local SVE of galaxies experience significant changes while the global SVE stays constant. For instance, we have found that recent fly-by interactions are able to generate two symmetric coherent spiral arms that leaves a clear imprint in the \sigmaZR maps. This is in contrast with the patterns seen for galaxies in which the spirals are not driven by flybys. Further, we found that the accretion of low mass satellites leave imprints in the \sigmaZR distributions that are not detected in the surface brightness maps.

Lastly, we found that the SVE properties vary for different stellar populations. In terms of the global SVE, \sigmaZ and \sigmaR are respectively 50\% lower and larger for the young and old populations compared to the velocity dispersion obtained with all the stellar particles. On the other hand, we showed that the global SVE of the young population is typically more oblate while the the old population present similar \sigmaZR values to the standard SVE. Moreover, the spatial distribution of the SVE is not the same for different populations and, in particular, the young population generally presents more complex patterns with spiral-like features in many of the galaxies. Future observations with new instruments like WEAVE \citep{2018SPIE10702E..1BD}, an integral field unit (IFU) that combines high spatial (1.5 arcsec per pixel) and spectral resolution (R$\approx$20000) will be able to examine these theoretical results.

In summary, our findings show that the global SVE of galaxies does not help to discriminate all the diverse mechanisms that affect the stars in the discs of galaxies. The spatial variations of the SVE are important because they can be used to reconstruct parts of the evolution history. The complexity of these results show the need of methods capable of revealing the local variations of the SVE in observations and simulations. Recently,  \cite{2017MNRAS.465.4956M} have been able to use a non-parametric approach with variable \sigmaZR across the disc of spiral galaxias obtaining similar results with more isotropic SVE inwards (\sigmaZR = 0.72) than in the outer parts (\sigmaZR = 0.30) for NGC1167. On the other hand, the  improved Schwarzschild’s orbit-superposition method from \cite{2020MNRAS.496.1579Z}  which allows to deproject the kinematic information without assuming a constant mass-to-light ratio across the galaxy. Thus, is is obvious that future studies should work towards the joint analysis from the theoretical and observational point of view.




\section*{Acknowledgements}
D. W-M and J. F-B acknowledge support through the RAVET project by the grant PID2019-107427GB-C32 from the Spanish Ministry of Science, Innovation and Universities (MCIU), and through the IAC project TRACES which is partially supported through the state budget and the regional budget of the Consejer\'ia de Econom\'ia, Industria, Comercio y Conocimiento of the Canary Islands Autonomous Community.


\section*{Data Availability}
The data underlying this article will be shared on reasonable request to the corresponding author.

\bibliographystyle{mnras}
\bibliography{SVE_Auriga} 

\begin{thebibliography}{}
\makeatletter
\relax
\def\mn@urlcharsother{\let\do\@makeother \do\$\do\&\do\#\do\^\do\_\do\%\do\~}
\def\mn@doi{\begingroup\mn@urlcharsother \@ifnextchar [ {\mn@doi@}
  {\mn@doi@[]}}
\def\mn@doi@[#1]#2{\def\@tempa{#1}\ifx\@tempa\@empty \href
  {http://dx.doi.org/#2} {doi:#2}\else \href {http://dx.doi.org/#2} {#1}\fi
  \endgroup}
\def\mn@eprint#1#2{\mn@eprint@#1:#2::\@nil}
\def\mn@eprint@arXiv#1{\href {http://arxiv.org/abs/#1} {{\tt arXiv:#1}}}
\def\mn@eprint@dblp#1{\href {http://dblp.uni-trier.de/rec/bibtex/#1.xml}
  {dblp:#1}}
\def\mn@eprint@#1:#2:#3:#4\@nil{\def\@tempa {#1}\def\@tempb {#2}\def\@tempc
  {#3}\ifx \@tempc \@empty \let \@tempc \@tempb \let \@tempb \@tempa \fi \ifx
  \@tempb \@empty \def\@tempb {arXiv}\fi \@ifundefined
  {mn@eprint@\@tempb}{\@tempb:\@tempc}{\expandafter \expandafter \csname
  mn@eprint@\@tempb\endcsname \expandafter{\@tempc}}}

\bibitem[\protect\citeauthoryear{{Ann} \& {Park}}{{Ann} \&
  {Park}}{2006}]{2006NewA...11..293A}
{Ann} H.~B.,  {Park} J.~C.,  2006, \mn@doi [\na]
  {10.1016/j.newast.2005.08.006}, \href
  {https://ui.adsabs.harvard.edu/abs/2006NewA...11..293A} {11, 293}

\bibitem[\protect\citeauthoryear{{Bittner}, {Gadotti}, {Elmegreen},
  {Athanassoula}, {Elmegreen}, {Bosma}  \& {Mu{\~n}oz-Mateos}}{{Bittner}
  et~al.}{2020a}]{2020IAUS..353..140B}
{Bittner} A.,  {Gadotti} D.~A.,  {Elmegreen} B.~G.,  {Athanassoula} E.,
  {Elmegreen} D.~M.,  {Bosma} A.,   {Mu{\~n}oz-Mateos} J.,  2020a, in {Valluri}
  M.,  {Sellwood} J.~A.,  eds,  IAU Symposium Vol. 353, IAU Symposium. pp
  140--143 (\mn@eprint {arXiv} {1910.01139}),
  \mn@doi{10.1017/S1743921319008160}

\bibitem[\protect\citeauthoryear{{Bittner} et~al.,}{{Bittner}
  et~al.}{2020b}]{2020A&A...643A..65B}
{Bittner} A.,  et~al., 2020b, \mn@doi [\aap] {10.1051/0004-6361/202038450},
  \href {https://ui.adsabs.harvard.edu/abs/2020A&A...643A..65B} {643, A65}

\bibitem[\protect\citeauthoryear{{Bl{\'a}zquez-Calero}
  et~al.,}{{Bl{\'a}zquez-Calero} et~al.}{2020}]{2020MNRAS.491.1800B}
{Bl{\'a}zquez-Calero} G.,  et~al., 2020, \mn@doi [\mnras]
  {10.1093/mnras/stz3125}, \href
  {https://ui.adsabs.harvard.edu/abs/2020MNRAS.491.1800B} {491, 1800}

\bibitem[\protect\citeauthoryear{{Buck}, {Obreja}, {Macci{\`o}}, {Minchev},
  {Dutton}  \& {Ostriker}}{{Buck} et~al.}{2020}]{2020MNRAS.491.3461B}
{Buck} T.,  {Obreja} A.,  {Macci{\`o}} A.~V.,  {Minchev} I.,  {Dutton} A.~A.,
  {Ostriker} J.~P.,  2020, \mn@doi [\mnras] {10.1093/mnras/stz3241}, \href
  {https://ui.adsabs.harvard.edu/abs/2020MNRAS.491.3461B} {491, 3461}

\bibitem[\protect\citeauthoryear{{B{\"u}denbender}, {van de Ven}  \&
  {Watkins}}{{B{\"u}denbender} et~al.}{2015}]{2015MNRAS.452..956B}
{B{\"u}denbender} A.,  {van de Ven} G.,   {Watkins} L.~L.,  2015, \mn@doi
  [\mnras] {10.1093/mnras/stv1314}, \href
  {https://ui.adsabs.harvard.edu/abs/2015MNRAS.452..956B} {452, 956}

\bibitem[\protect\citeauthoryear{{Burstein}}{{Burstein}}{1979}]{1979ApJ...234..829B}
{Burstein} D.,  1979, \mn@doi [\apj] {10.1086/157563}, \href
  {https://ui.adsabs.harvard.edu/abs/1979ApJ...234..829B} {234, 829}

\bibitem[\protect\citeauthoryear{{Cappellari} \& {Copin}}{{Cappellari} \&
  {Copin}}{2003}]{2003MNRAS.342..345C}
{Cappellari} M.,  {Copin} Y.,  2003, \mn@doi [\mnras]
  {10.1046/j.1365-8711.2003.06541.x}, \href
  {https://ui.adsabs.harvard.edu/abs/2003MNRAS.342..345C} {342, 345}

\bibitem[\protect\citeauthoryear{{Cappellari} et~al.,}{{Cappellari}
  et~al.}{2007}]{2007MNRAS.379..418C}
{Cappellari} M.,  et~al., 2007, \mn@doi [\mnras]
  {10.1111/j.1365-2966.2007.11963.x}, \href
  {https://ui.adsabs.harvard.edu/abs/2007MNRAS.379..418C} {379, 418}

\bibitem[\protect\citeauthoryear{{Cappellari} et~al.,}{{Cappellari}
  et~al.}{2008}]{2008IAUS..245..215C}
{Cappellari} M.,  et~al., 2008, \mn@doi [\mnras] {10.1017/S1743921308017687},
  \href {https://ui.adsabs.harvard.edu/abs/2008IAUS..245..215C} {245, 215}

\bibitem[\protect\citeauthoryear{{Courteau} et~al.,}{{Courteau}
  et~al.}{2014}]{2014RvMP...86...47C}
{Courteau} S.,  et~al., 2014, \mn@doi [Reviews of Modern Physics]
  {10.1103/RevModPhys.86.47}, \href
  {https://ui.adsabs.harvard.edu/abs/2014RvMP...86...47C} {86, 47}

\bibitem[\protect\citeauthoryear{{Dalton} et~al.,}{{Dalton}
  et~al.}{2018}]{2018SPIE10702E..1BD}
{Dalton} G.,  et~al., 2018, in {Evans} C.~J.,  {Simard} L.,   {Takami} H.,
  eds,  Society of Photo-Optical Instrumentation Engineers (SPIE) Conference
  Series Vol. 10702, Ground-based and Airborne Instrumentation for Astronomy
  VII. p. 107021B, \mn@doi{10.1117/12.2312031}

\bibitem[\protect\citeauthoryear{{Danver}}{{Danver}}{1942}]{1942AnLun..10..115D}
{Danver} C.-G.,  1942, Annals of the Observatory of Lund, \href
  {https://ui.adsabs.harvard.edu/abs/1942AnLun..10..115D} {10, 115}

\bibitem[\protect\citeauthoryear{{Davis}, {Efstathiou}, {Frenk}  \&
  {White}}{{Davis} et~al.}{1985}]{1985ApJ...292..371D}
{Davis} M.,  {Efstathiou} G.,  {Frenk} C.~S.,   {White} S.~D.~M.,  1985,
  \mn@doi [\apj] {10.1086/163168}, \href
  {https://ui.adsabs.harvard.edu/abs/1985ApJ...292..371D} {292, 371}

\bibitem[\protect\citeauthoryear{{Emsellem}, {Dejonghe}  \& {Bacon}}{{Emsellem}
  et~al.}{1999}]{1999MNRAS.303..495E}
{Emsellem} E.,  {Dejonghe} H.,   {Bacon} R.,  1999, \mn@doi [\mnras]
  {10.1046/j.1365-8711.1999.02210.x}, \href
  {https://ui.adsabs.harvard.edu/abs/1999MNRAS.303..495E} {303, 495}

\bibitem[\protect\citeauthoryear{{Emsellem} et~al.,}{{Emsellem}
  et~al.}{2011}]{2011MNRAS.414..888E}
{Emsellem} E.,  et~al., 2011, \mn@doi [\mnras]
  {10.1111/j.1365-2966.2011.18496.x}, \href
  {https://ui.adsabs.harvard.edu/abs/2011MNRAS.414..888E} {414, 888}

\bibitem[\protect\citeauthoryear{{Everall}, {Evans}, {Belokurov}  \&
  {Sch{\"o}nrich}}{{Everall} et~al.}{2019}]{2019MNRAS.489..910E}
{Everall} A.,  {Evans} N.~W.,  {Belokurov} V.,   {Sch{\"o}nrich} R.,  2019,
  \mn@doi [\mnras] {10.1093/mnras/stz2217}, \href
  {https://ui.adsabs.harvard.edu/abs/2019MNRAS.489..910E} {489, 910}

\bibitem[\protect\citeauthoryear{{Falcon-Barroso} \& {Martig}}{{Falcon-Barroso}
  \& {Martig}}{2020}]{2020arXiv201112023F}
{Falcon-Barroso} J.,  {Martig} M.,  2020, arXiv e-prints, \href
  {https://ui.adsabs.harvard.edu/abs/2020arXiv201112023F} {p. arXiv:2011.12023}

\bibitem[\protect\citeauthoryear{{Few} \& {Madore}}{{Few} \&
  {Madore}}{1986}]{1986MNRAS.222..673F}
{Few} J. M.~A.,  {Madore} B.~F.,  1986, \mn@doi [\mnras]
  {10.1093/mnras/222.4.673}, \href
  {https://ui.adsabs.harvard.edu/abs/1986MNRAS.222..673F} {222, 673}

\bibitem[\protect\citeauthoryear{{F{\"o}rster Schreiber} et~al.,}{{F{\"o}rster
  Schreiber} et~al.}{2009}]{2009ApJ...706.1364F}
{F{\"o}rster Schreiber} N.~M.,  et~al., 2009, \mn@doi [\apj]
  {10.1088/0004-637X/706/2/1364}, \href
  {https://ui.adsabs.harvard.edu/abs/2009ApJ...706.1364F} {706, 1364}

\bibitem[\protect\citeauthoryear{{Gadotti} et~al.,}{{Gadotti}
  et~al.}{2020}]{2020A&A...643A..14G}
{Gadotti} D.~A.,  et~al., 2020, \mn@doi [\aap] {10.1051/0004-6361/202038448},
  \href {https://ui.adsabs.harvard.edu/abs/2020A&A...643A..14G} {643, A14}

\bibitem[\protect\citeauthoryear{{Gaia Collaboration} et~al.,}{{Gaia
  Collaboration} et~al.}{2016}]{2016A&A...595A...1G}
{Gaia Collaboration} et~al., 2016, \mn@doi [\aap]
  {10.1051/0004-6361/201629272}, \href
  {https://ui.adsabs.harvard.edu/abs/2016A&A...595A...1G} {595, A1}

\bibitem[\protect\citeauthoryear{{Gargiulo} et~al.,}{{Gargiulo}
  et~al.}{2019}]{2019MNRAS.489.5742G}
{Gargiulo} I.~D.,  et~al., 2019, \mn@doi [\mnras] {10.1093/mnras/stz2536},
  \href {https://ui.adsabs.harvard.edu/abs/2019MNRAS.489.5742G} {489, 5742}

\bibitem[\protect\citeauthoryear{{Gerssen} \& {Shapiro Griffin}}{{Gerssen} \&
  {Shapiro Griffin}}{2012}]{2012MNRAS.423.2726G}
{Gerssen} J.,  {Shapiro Griffin} K.,  2012, \mn@doi [\mnras]
  {10.1111/j.1365-2966.2012.21078.x}, \href
  {https://ui.adsabs.harvard.edu/abs/2012MNRAS.423.2726G} {423, 2726}

\bibitem[\protect\citeauthoryear{{G{\'o}mez}, {White}, {Grand}, {Marinacci},
  {Springel}  \& {Pakmor}}{{G{\'o}mez} et~al.}{2017}]{2017MNRAS.465.3446G}
{G{\'o}mez} F.~A.,  {White} S. D.~M.,  {Grand} R. J.~J.,  {Marinacci} F.,
  {Springel} V.,   {Pakmor} R.,  2017, \mn@doi [\mnras]
  {10.1093/mnras/stw2957}, \href
  {https://ui.adsabs.harvard.edu/abs/2017MNRAS.465.3446G} {465, 3446}

\bibitem[\protect\citeauthoryear{{Grand}, {Springel}, {G{\'o}mez}, {Marinacci},
  {Pakmor}, {Campbell}  \& {Jenkins}}{{Grand}
  et~al.}{2016a}]{2016MNRAS.459..199G}
{Grand} R. J.~J.,  {Springel} V.,  {G{\'o}mez} F.~A.,  {Marinacci} F.,
  {Pakmor} R.,  {Campbell} D. J.~R.,   {Jenkins} A.,  2016a, \mn@doi [\mnras]
  {10.1093/mnras/stw601}, \href
  {https://ui.adsabs.harvard.edu/abs/2016MNRAS.459..199G} {459, 199}

\bibitem[\protect\citeauthoryear{{Grand} et~al.,}{{Grand}
  et~al.}{2016b}]{2016MNRAS.460L..94G}
{Grand} R. J.~J.,  et~al., 2016b, \mn@doi [\mnras] {10.1093/mnrasl/slw086},
  \href {https://ui.adsabs.harvard.edu/abs/2016MNRAS.460L..94G} {460, L94}

\bibitem[\protect\citeauthoryear{{Grand} et~al.,}{{Grand}
  et~al.}{2017}]{AurigaProject}
{Grand} R. J.~J.,  et~al., 2017, \mn@doi [\mnras] {10.1093/mnras/stx071}, \href
  {https://ui.adsabs.harvard.edu/abs/2017MNRAS.467..179G} {467, 179}

\bibitem[\protect\citeauthoryear{{Green} et~al.,}{{Green}
  et~al.}{2014}]{2014MNRAS.437.1070G}
{Green} A.~W.,  et~al., 2014, \mn@doi [\mnras] {10.1093/mnras/stt1882}, \href
  {https://ui.adsabs.harvard.edu/abs/2014MNRAS.437.1070G} {437, 1070}

\bibitem[\protect\citeauthoryear{{Hagen}, {Helmi}, {de Zeeuw}  \&
  {Posti}}{{Hagen} et~al.}{2019}]{2019A&A...629A..70H}
{Hagen} J.~H.~J.,  {Helmi} A.,  {de Zeeuw} P.~T.,   {Posti} L.,  2019, \mn@doi
  [\aap] {10.1051/0004-6361/201935264}, \href
  {https://ui.adsabs.harvard.edu/abs/2019A&A...629A..70H} {629, A70}

\bibitem[\protect\citeauthoryear{{Kalinova} et~al.,}{{Kalinova}
  et~al.}{2017}]{2017MNRAS.469.2539K}
{Kalinova} V.,  et~al., 2017, \mn@doi [\mnras] {10.1093/mnras/stx901}, \href
  {https://ui.adsabs.harvard.edu/abs/2017MNRAS.469.2539K} {469, 2539}

\bibitem[\protect\citeauthoryear{{Kipper}, {Tenjes}, {Tihhonova}, {Tamm}  \&
  {Tempel}}{{Kipper} et~al.}{2016}]{2016MNRAS.460.2720K}
{Kipper} R.,  {Tenjes} P.,  {Tihhonova} O.,  {Tamm} A.,   {Tempel} E.,  2016,
  \mn@doi [\mnras] {10.1093/mnras/stw1194}, \href
  {https://ui.adsabs.harvard.edu/abs/2016MNRAS.460.2720K} {460, 2720}

\bibitem[\protect\citeauthoryear{{Kormendy} \& {Ho}}{{Kormendy} \&
  {Ho}}{2013}]{2013ARA&A..51..511K}
{Kormendy} J.,  {Ho} L.~C.,  2013, \mn@doi [\araa]
  {10.1146/annurev-astro-082708-101811}, \href
  {https://ui.adsabs.harvard.edu/abs/2013ARA&A..51..511K} {51, 511}

\bibitem[\protect\citeauthoryear{{Krajnovi{\'c}} et~al.,}{{Krajnovi{\'c}}
  et~al.}{2018}]{2018MNRAS.477.3030K}
{Krajnovi{\'c}} D.,  et~al., 2018, \mn@doi [\mnras] {10.1093/mnras/sty778},
  \href {https://ui.adsabs.harvard.edu/abs/2018MNRAS.477.3030K} {477, 3030}

\bibitem[\protect\citeauthoryear{{Mackereth} et~al.,}{{Mackereth}
  et~al.}{2019}]{2019MNRAS.489..176M}
{Mackereth} J.~T.,  et~al., 2019, \mn@doi [\mnras] {10.1093/mnras/stz1521},
  \href {https://ui.adsabs.harvard.edu/abs/2019MNRAS.489..176M} {489, 176}

\bibitem[\protect\citeauthoryear{{Majewski} et~al.,}{{Majewski}
  et~al.}{2017}]{2017AJ....154...94M}
{Majewski} S.~R.,  et~al., 2017, \mn@doi [\aj] {10.3847/1538-3881/aa784d},
  \href {https://ui.adsabs.harvard.edu/abs/2017AJ....154...94M} {154, 94}

\bibitem[\protect\citeauthoryear{{Marchuk} \& {Sotnikova}}{{Marchuk} \&
  {Sotnikova}}{2017}]{2017MNRAS.465.4956M}
{Marchuk} A.~A.,  {Sotnikova} N.~Y.,  2017, \mn@doi [\mnras]
  {10.1093/mnras/stw3092}, \href
  {https://ui.adsabs.harvard.edu/abs/2017MNRAS.465.4956M} {465, 4956}

\bibitem[\protect\citeauthoryear{{Marinacci}, {Pakmor}  \&
  {Springel}}{{Marinacci} et~al.}{2014a}]{2014MNRAS.437.1750M}
{Marinacci} F.,  {Pakmor} R.,   {Springel} V.,  2014a, \mn@doi [\mnras]
  {10.1093/mnras/stt2003}, \href
  {https://ui.adsabs.harvard.edu/abs/2014MNRAS.437.1750M} {437, 1750}

\bibitem[\protect\citeauthoryear{{Marinacci}, {Pakmor}, {Springel}  \&
  {Simpson}}{{Marinacci} et~al.}{2014b}]{2014MNRAS.442.3745M}
{Marinacci} F.,  {Pakmor} R.,  {Springel} V.,   {Simpson} C.~M.,  2014b,
  \mn@doi [\mnras] {10.1093/mnras/stu1136}, \href
  {https://ui.adsabs.harvard.edu/abs/2014MNRAS.442.3745M} {442, 3745}

\bibitem[\protect\citeauthoryear{{Martig}, {Bournaud}, {Croton}, {Dekel}  \&
  {Teyssier}}{{Martig} et~al.}{2012}]{2012ApJ...756...26M}
{Martig} M.,  {Bournaud} F.,  {Croton} D.~J.,  {Dekel} A.,   {Teyssier} R.,
  2012, \mn@doi [\apj] {10.1088/0004-637X/756/1/26}, \href
  {https://ui.adsabs.harvard.edu/abs/2012ApJ...756...26M} {756, 26}

\bibitem[\protect\citeauthoryear{{Mart{\'\i}n-Navarro}
  et~al.,}{{Mart{\'\i}n-Navarro} et~al.}{2012}]{2012MNRAS.427.1102M}
{Mart{\'\i}n-Navarro} I.,  et~al., 2012, \mn@doi [\mnras]
  {10.1111/j.1365-2966.2012.21929.x}, \href
  {https://ui.adsabs.harvard.edu/abs/2012MNRAS.427.1102M} {427, 1102}

\bibitem[\protect\citeauthoryear{{Martinez-Valpuesta}, {Aguerri},
  {Gonz{\'a}lez-Garc{\'\i}a}, {Dalla Vecchia}  \&
  {Stringer}}{{Martinez-Valpuesta} et~al.}{2017}]{2017MNRAS.464.1502M}
{Martinez-Valpuesta} I.,  {Aguerri} J. A.~L.,  {Gonz{\'a}lez-Garc{\'\i}a}
  A.~C.,  {Dalla Vecchia} C.,   {Stringer} M.,  2017, \mn@doi [\mnras]
  {10.1093/mnras/stw2500}, \href
  {https://ui.adsabs.harvard.edu/abs/2017MNRAS.464.1502M} {464, 1502}

\bibitem[\protect\citeauthoryear{{Martinsson}, {Verheijen}, {Westfall},
  {Bershady}, {Schechtman-Rook}, {Andersen}  \& {Swaters}}{{Martinsson}
  et~al.}{2013}]{2013A&A...557A.130M}
{Martinsson} T. P.~K.,  {Verheijen} M. A.~W.,  {Westfall} K.~B.,  {Bershady}
  M.~A.,  {Schechtman-Rook} A.,  {Andersen} D.~R.,   {Swaters} R.~A.,  2013,
  \mn@doi [\aap] {10.1051/0004-6361/201220515}, \href
  {https://ui.adsabs.harvard.edu/abs/2013A&A...557A.130M} {557, A130}

\bibitem[\protect\citeauthoryear{{Mogotsi} \& {Romeo}}{{Mogotsi} \&
  {Romeo}}{2019}]{2019MNRAS.489.3797M}
{Mogotsi} K.~M.,  {Romeo} A.~B.,  2019, \mn@doi [\mnras]
  {10.1093/mnras/stz2370}, \href
  {https://ui.adsabs.harvard.edu/abs/2019MNRAS.489.3797M} {489, 3797}

\bibitem[\protect\citeauthoryear{{Pillepich} et~al.,}{{Pillepich}
  et~al.}{2018}]{2018MNRAS.473.4077P}
{Pillepich} A.,  et~al., 2018, \mn@doi [\mnras] {10.1093/mnras/stx2656}, \href
  {https://ui.adsabs.harvard.edu/abs/2018MNRAS.473.4077P} {473, 4077}

\bibitem[\protect\citeauthoryear{{Pillepich} et~al.,}{{Pillepich}
  et~al.}{2019}]{2019MNRAS.490.3196P}
{Pillepich} A.,  et~al., 2019, \mn@doi [\mnras] {10.1093/mnras/stz2338}, \href
  {https://ui.adsabs.harvard.edu/abs/2019MNRAS.490.3196P} {490, 3196}

\bibitem[\protect\citeauthoryear{{Pinna}, {Falc{\'o}n-Barroso}, {Martig},
  {Mart{\'\i}nez-Valpuesta}, {M{\'e}ndez-Abreu}, {van de Ven}, {Leaman}  \&
  {Lyubenova}}{{Pinna} et~al.}{2018}]{PinnaSVE}
{Pinna} F.,  {Falc{\'o}n-Barroso} J.,  {Martig} M.,  {Mart{\'\i}nez-Valpuesta}
  I.,  {M{\'e}ndez-Abreu} J.,  {van de Ven} G.,  {Leaman} R.,   {Lyubenova} M.,
   2018, \mn@doi [\mnras] {10.1093/mnras/stx3331}, \href
  {https://ui.adsabs.harvard.edu/abs/2018MNRAS.475.2697P} {475, 2697}

\bibitem[\protect\citeauthoryear{{Pinna} et~al.,}{{Pinna}
  et~al.}{2019}]{2019A&A...625A..95P}
{Pinna} F.,  et~al., 2019, \mn@doi [\aap] {10.1051/0004-6361/201935154}, \href
  {https://ui.adsabs.harvard.edu/abs/2019A&A...625A..95P} {625, A95}

\bibitem[\protect\citeauthoryear{{Planck Collaboration} et~al.,}{{Planck
  Collaboration} et~al.}{2014}]{2014A&A...571A..16P}
{Planck Collaboration} et~al., 2014, \mn@doi [\aap]
  {10.1051/0004-6361/201321591}, \href
  {https://ui.adsabs.harvard.edu/abs/2014A&A...571A..16P} {571, A16}

\bibitem[\protect\citeauthoryear{{Pohlen}, {Beckman}, {H{\"u}ttemeister},
  {Knapen}, {Erwin}  \& {Dettmar}}{{Pohlen} et~al.}{2004}]{2004ASSL..319..713P}
{Pohlen} M.,  {Beckman} J.~E.,  {H{\"u}ttemeister} S.,  {Knapen} J.~H.,
  {Erwin} P.,   {Dettmar} R.~J.,  2004, {Stellar Disk Truncations: Where do we
  stand?}.
p.~713, \mn@doi{10.1007/978-1-4020-2862-5_61}

\bibitem[\protect\citeauthoryear{{Reshetnikov} \& {Combes}}{{Reshetnikov} \&
  {Combes}}{1998}]{1998A&A...337....9R}
{Reshetnikov} V.,  {Combes} F.,  1998, \aap, \href
  {https://ui.adsabs.harvard.edu/abs/1998A&A...337....9R} {337, 9}

\bibitem[\protect\citeauthoryear{{Reshetnikov}, {Mosenkov}, {Moiseev}, {Kotov}
  \& {Savchenko}}{{Reshetnikov} et~al.}{2016}]{2016MNRAS.461.4233R}
{Reshetnikov} V.~P.,  {Mosenkov} A.~V.,  {Moiseev} A.~V.,  {Kotov} S.~S.,
  {Savchenko} S.~S.,  2016, \mn@doi [\mnras] {10.1093/mnras/stw1623}, \href
  {https://ui.adsabs.harvard.edu/abs/2016MNRAS.461.4233R} {461, 4233}

\bibitem[\protect\citeauthoryear{{Rodionov} \& {Sotnikova}}{{Rodionov} \&
  {Sotnikova}}{2013}]{2013MNRAS.434.2373R}
{Rodionov} S.~A.,  {Sotnikova} N.~Y.,  2013, \mn@doi [\mnras]
  {10.1093/mnras/stt1183}, \href
  {https://ui.adsabs.harvard.edu/abs/2013MNRAS.434.2373R} {434, 2373}

\bibitem[\protect\citeauthoryear{{Rosado-Belza} et~al.,}{{Rosado-Belza}
  et~al.}{2020}]{2020A&A...644A.116R}
{Rosado-Belza} D.,  et~al., 2020, \mn@doi [\aap] {10.1051/0004-6361/202039530},
  \href {https://ui.adsabs.harvard.edu/abs/2020A&A...644A.116R} {644, A116}

\bibitem[\protect\citeauthoryear{{S{\'a}nchez-Menguiano}
  et~al.,}{{S{\'a}nchez-Menguiano} et~al.}{2016}]{2016ApJ...830L..40S}
{S{\'a}nchez-Menguiano} L.,  et~al., 2016, \mn@doi [\apjl]
  {10.3847/2041-8205/830/2/L40}, \href
  {https://ui.adsabs.harvard.edu/abs/2016ApJ...830L..40S} {830, L40}

\bibitem[\protect\citeauthoryear{{S{\'a}nchez} et~al.,}{{S{\'a}nchez}
  et~al.}{2012}]{2012A&A...538A...8S}
{S{\'a}nchez} S.~F.,  et~al., 2012, \mn@doi [\aap]
  {10.1051/0004-6361/201117353}, \href
  {https://ui.adsabs.harvard.edu/abs/2012A&A...538A...8S} {538, A8}

\bibitem[\protect\citeauthoryear{{Schaye} et~al.,}{{Schaye}
  et~al.}{2015}]{2015MNRAS.446..521S}
{Schaye} J.,  et~al., 2015, \mn@doi [\mnras] {10.1093/mnras/stu2058}, \href
  {https://ui.adsabs.harvard.edu/abs/2015MNRAS.446..521S} {446, 521}

\bibitem[\protect\citeauthoryear{{Schulze}, {Remus}, {Dolag}, {Burkert},
  {Emsellem}  \& {van de Ven}}{{Schulze} et~al.}{2018}]{2018MNRAS.480.4636S}
{Schulze} F.,  {Remus} R.-S.,  {Dolag} K.,  {Burkert} A.,  {Emsellem} E.,
  {van de Ven} G.,  2018, \mn@doi [\mnras] {10.1093/mnras/sty2090}, \href
  {https://ui.adsabs.harvard.edu/abs/2018MNRAS.480.4636S} {480, 4636}

\bibitem[\protect\citeauthoryear{{Sellwood}}{{Sellwood}}{2013}]{2013ApJ...769L..24S}
{Sellwood} J.~A.,  2013, \mn@doi [\apjl] {10.1088/2041-8205/769/2/L24}, \href
  {https://ui.adsabs.harvard.edu/abs/2013ApJ...769L..24S} {769, L24}

\bibitem[\protect\citeauthoryear{{Shapiro}, {Gerssen}  \& {van der
  Marel}}{{Shapiro} et~al.}{2003}]{2003AJ....126.2707S}
{Shapiro} K.~L.,  {Gerssen} J.,   {van der Marel} R.~P.,  2003, \mn@doi [\aj]
  {10.1086/379306}, \href
  {https://ui.adsabs.harvard.edu/abs/2003AJ....126.2707S} {126, 2707}

\bibitem[\protect\citeauthoryear{{Shlosman}, {Frank}  \& {Begelman}}{{Shlosman}
  et~al.}{1989}]{1989Natur.338...45S}
{Shlosman} I.,  {Frank} J.,   {Begelman} M.~C.,  1989, \mn@doi [\nat]
  {10.1038/338045a0}, \href
  {https://ui.adsabs.harvard.edu/abs/1989Natur.338...45S} {338, 45}

\bibitem[\protect\citeauthoryear{{Simion}, {Shen}, {Koposov}, {Ness},
  {Freeman}, {Bland-Hawthorn}  \& {Lewis}}{{Simion}
  et~al.}{2020}]{2020arXiv201113905S}
{Simion} I.~T.,  {Shen} J.,  {Koposov} S.~E.,  {Ness} M.,  {Freeman} K.,
  {Bland-Hawthorn} J.,   {Lewis} G.~F.,  2020, arXiv e-prints, \href
  {https://ui.adsabs.harvard.edu/abs/2020arXiv201113905S} {p. arXiv:2011.13905}

\bibitem[\protect\citeauthoryear{{Soto}, {Rich}  \& {Kuijken}}{{Soto}
  et~al.}{2007}]{2007ApJ...665L..31S}
{Soto} M.,  {Rich} R.~M.,   {Kuijken} K.,  2007, \mn@doi [\apjl]
  {10.1086/521098}, \href
  {https://ui.adsabs.harvard.edu/abs/2007ApJ...665L..31S} {665, L31}

\bibitem[\protect\citeauthoryear{{Springel}}{{Springel}}{2010}]{2010ARA&A..48..391S}
{Springel} V.,  2010, \mn@doi [\araa] {10.1146/annurev-astro-081309-130914},
  \href {https://ui.adsabs.harvard.edu/abs/2010ARA&A..48..391S} {48, 391}

\bibitem[\protect\citeauthoryear{{Springel}, {White}, {Tormen}  \&
  {Kauffmann}}{{Springel} et~al.}{2001}]{2001MNRAS.328..726S}
{Springel} V.,  {White} S. D.~M.,  {Tormen} G.,   {Kauffmann} G.,  2001,
  \mn@doi [\mnras] {10.1046/j.1365-8711.2001.04912.x}, \href
  {https://ui.adsabs.harvard.edu/abs/2001MNRAS.328..726S} {328, 726}

\bibitem[\protect\citeauthoryear{{Springel} et~al.,}{{Springel}
  et~al.}{2005}]{2005Natur.435..629S}
{Springel} V.,  et~al., 2005, \mn@doi [\nat] {10.1038/nature03597}, \href
  {https://ui.adsabs.harvard.edu/abs/2005Natur.435..629S} {435, 629}

\bibitem[\protect\citeauthoryear{{Tempel} \& {Tenjes}}{{Tempel} \&
  {Tenjes}}{2006}]{2006MNRAS.371.1269T}
{Tempel} E.,  {Tenjes} P.,  2006, \mn@doi [\mnras]
  {10.1111/j.1365-2966.2006.10741.x}, \href
  {https://ui.adsabs.harvard.edu/abs/2006MNRAS.371.1269T} {371, 1269}

\bibitem[\protect\citeauthoryear{{Toomre}}{{Toomre}}{1964}]{1964ApJ...139.1217T}
{Toomre} A.,  1964, \mn@doi [\apj] {10.1086/147861}, \href
  {https://ui.adsabs.harvard.edu/abs/1964ApJ...139.1217T} {139, 1217}

\bibitem[\protect\citeauthoryear{{Tsikoudi}}{{Tsikoudi}}{1979}]{1979ApJ...234..842T}
{Tsikoudi} V.,  1979, \mn@doi [\apj] {10.1086/157565}, \href
  {https://ui.adsabs.harvard.edu/abs/1979ApJ...234..842T} {234, 842}

\bibitem[\protect\citeauthoryear{{Vasiliev} \& {Valluri}}{{Vasiliev} \&
  {Valluri}}{2020}]{2020ApJ...889...39V}
{Vasiliev} E.,  {Valluri} M.,  2020, \mn@doi [\apj] {10.3847/1538-4357/ab5fe0},
  \href {https://ui.adsabs.harvard.edu/abs/2020ApJ...889...39V} {889, 39}

\bibitem[\protect\citeauthoryear{{Walo-Mart{\'\i}n}, {Falc{\'o}n-Barroso},
  {Dalla Vecchia}, {P{\'e}rez}  \& {Negri}}{{Walo-Mart{\'\i}n}
  et~al.}{2020}]{2020MNRAS.494.5652W}
{Walo-Mart{\'\i}n} D.,  {Falc{\'o}n-Barroso} J.,  {Dalla Vecchia} C.,
  {P{\'e}rez} I.,   {Negri} A.,  2020, \mn@doi [\mnras]
  {10.1093/mnras/staa1066}, \href
  {https://ui.adsabs.harvard.edu/abs/2020MNRAS.494.5652W} {494, 5652}

\bibitem[\protect\citeauthoryear{{Wisnioski} et~al.,}{{Wisnioski}
  et~al.}{2015}]{2015ApJ...799..209W}
{Wisnioski} E.,  et~al., 2015, \mn@doi [\apj] {10.1088/0004-637X/799/2/209},
  \href {https://ui.adsabs.harvard.edu/abs/2015ApJ...799..209W} {799, 209}

\bibitem[\protect\citeauthoryear{{Zhao}, {Spergel}  \& {Rich}}{{Zhao}
  et~al.}{1994}]{1994AJ....108.2154Z}
{Zhao} H.,  {Spergel} D.~N.,   {Rich} R.~M.,  1994, \mn@doi [\aj]
  {10.1086/117227}, \href
  {https://ui.adsabs.harvard.edu/abs/1994AJ....108.2154Z} {108, 2154}

\bibitem[\protect\citeauthoryear{{Zhu} et~al.,}{{Zhu}
  et~al.}{2020}]{2020MNRAS.496.1579Z}
{Zhu} L.,  et~al., 2020, \mn@doi [\mnras] {10.1093/mnras/staa1584}, \href
  {https://ui.adsabs.harvard.edu/abs/2020MNRAS.496.1579Z} {496, 1579}

\bibitem[\protect\citeauthoryear{{de Lorenzo-C{\'a}ceres}, {M{\'e}ndez-Abreu},
  {Thorne}  \& {Costantin}}{{de Lorenzo-C{\'a}ceres}
  et~al.}{2019}]{2019MNRAS.484..665D}
{de Lorenzo-C{\'a}ceres} A.,  {M{\'e}ndez-Abreu} J.,  {Thorne} B.,
  {Costantin} L.,  2019, \mn@doi [\mnras] {10.1093/mnras/sty3520}, \href
  {https://ui.adsabs.harvard.edu/abs/2019MNRAS.484..665D} {484, 665}

\bibitem[\protect\citeauthoryear{{van de Sande} et~al.,}{{van de Sande}
  et~al.}{2017}]{2017ApJ...835..104V}
{van de Sande} J.,  et~al., 2017, \mn@doi [\apj] {10.3847/1538-4357/835/1/104},
  \href {https://ui.adsabs.harvard.edu/abs/2017ApJ...835..104V} {835, 104}

\bibitem[\protect\citeauthoryear{{van de Ven}, {de Zeeuw}  \& {van den
  Bosch}}{{van de Ven} et~al.}{2008}]{2008MNRAS.385..614V}
{van de Ven} G.,  {de Zeeuw} P.~T.,   {van den Bosch} R.~C.~E.,  2008, \mn@doi
  [\mnras] {10.1111/j.1365-2966.2008.12873.x}, \href
  {https://ui.adsabs.harvard.edu/abs/2008MNRAS.385..614V} {385, 614}

\bibitem[\protect\citeauthoryear{{van den Bosch}, {van de Ven}, {Verolme},
  {Cappellari}  \& {de Zeeuw}}{{van den Bosch}
  et~al.}{2008}]{2008MNRAS.385..647V}
{van den Bosch} R.~C.~E.,  {van de Ven} G.,  {Verolme} E.~K.,  {Cappellari} M.,
    {de Zeeuw} P.~T.,  2008, \mn@doi [\mnras]
  {10.1111/j.1365-2966.2008.12874.x}, \href
  {https://ui.adsabs.harvard.edu/abs/2008MNRAS.385..647V} {385, 647}

\bibitem[\protect\citeauthoryear{{van der Kruit}}{{van der
  Kruit}}{1979}]{1979A&AS...38...15V}
{van der Kruit} P.~C.,  1979, \aaps, \href
  {https://ui.adsabs.harvard.edu/abs/1979A&AS...38...15V} {38, 15}

\bibitem[\protect\citeauthoryear{{van der Kruit}}{{van der
  Kruit}}{1988}]{1988A&A...192..117V}
{van der Kruit} P.~C.,  1988, \aap, \href
  {https://ui.adsabs.harvard.edu/abs/1988A&A...192..117V} {192, 117}

\bibitem[\protect\citeauthoryear{{van der Kruit} \& {Searle}}{{van der Kruit}
  \& {Searle}}{1981}]{1981A&A....95..105V}
{van der Kruit} P.~C.,  {Searle} L.,  1981, \aap, \href
  {https://ui.adsabs.harvard.edu/abs/1981A&A....95..105V} {95, 105}

\bibitem[\protect\citeauthoryear{{van der Kruit} \& {de Grijs}}{{van der Kruit}
  \& {de Grijs}}{1999}]{1999A&A...352..129V}
{van der Kruit} P.~C.,  {de Grijs} R.,  1999, \aap, \href
  {https://ui.adsabs.harvard.edu/abs/1999A&A...352..129V} {352, 129}

\bibitem[\protect\citeauthoryear{{van der Marel} \& {Franx}}{{van der Marel} \&
  {Franx}}{1993}]{1993ApJ...407..525V}
{van der Marel} R.~P.,  {Franx} M.,  1993, \mn@doi [\apj] {10.1086/172534},
  \href {https://ui.adsabs.harvard.edu/abs/1993ApJ...407..525V} {407, 525}

\makeatother
\end{thebibliography}



\bsp	
\label{lastpage}
\end{document}